\documentclass{aa}
\usepackage{amsmath}
\usepackage{graphicx}
\usepackage{natbib}
\bibpunct{(}{)}{;}{a}{}{,}
\usepackage{txfonts}
\newcommand{\cmt}{\,cm$^{-3}$}
\newcommand{\cmd}{\,cm$^{-2}$}
\newcommand{\kms}{\,km\,s$^{-1}$}
\newcommand{\myr}{\,$M_{\sun}\,{\rm yr}^{-1}$}
\newcommand{\ha}{H$\alpha$}
\newcommand{\hb}{H$\beta$}
\newcommand{\ro}{\,$R_{\sun}$}
\newcommand{\mo}{\,$M_{\sun}$}
\newcommand{\lo}{\,$L_{\sun}$}
\begin{document}
%
\title{Ionization structure of hot components in symbiotic binaries 
       during active phases}

\author{Z. Carikov\'{a} \and A. Skopal}
\institute{Astronomical Institute, Slovak Academy of Sciences,
        059~60 Tatransk\'{a} Lomnica, Slovakia\\
		\email{zcarikova@ta3.sk}\\
		\email{skopal@ta3.sk}}
		
\date{Received / Accepted}
\abstract
{
During active phases of symbiotic binaries, an optically thick 
medium in the form of a flared disk develops around their hot 
stars. During quiescent phases, this structure is not evident. 
}
{
We propose the formation of a flared neutral disk-like structure 
around the rotating white dwarf (WD) in symbiotic binaries. 
}
{
We applied the wind compression model and calculated the ionization 
boundaries in the compressed wind from the WD using the equation of 
photoionization equilibrium.
}
{
During active phases, the compression of the enhanced wind from 
the rotating WD can form a neutral disk-like zone at the 
equatorial plane, while the remainder of the sphere above/below 
the disk is ionized. 
The hydrogen column density throughout the neutral zone and the 
emission measure of the ionized fraction of the wind, calculated 
for the mass loss rate from the WD, $\dot{M} = 2\times10^{-6}$\myr\ 
with $v_{\infty} = 2\,000$\kms, are consistent with those derived 
from observations. During quiescent phases, the neutral disk-like 
structure cannot be created because of insufficient mass loss 
rate. 
}
{
Formation of the neutral disk-like zone at the equatorial plane 
is connected with the enhanced wind from the rotating WD,
observed during active phases of symbiotic binaries.
This probably represents a common origin of warm
pseudophotospheres, indicated in the spectrum of active
symbiotic binaries with a high orbital inclination. 
}

\maketitle
\section{Introduction}
Symbiotic stars are interacting binary systems comprising a cool 
giant as the donor star and a compact star, mostly a white dwarf
(WD), as the accretor. This composition constrains large orbital 
periods, which are typically a few years, but can be even 
longer. Principal process of the interaction is the mass transfer 
between the binary components via the stellar wind from the giant. 
The following accretion heats up the WD to 
$\ga 10^5$\,K and increases its luminosity to 
$\sim 10^{2}-10^{4}$\lo. Such a hot and luminous WD is capable 
of ionizing neutral wind particles of both the stars, giving rise 
to strong nebular radiation. The spectrum of symbiotic stars 
thus consists of three main components of radiation, two stellar 
and one nebular. 
Many aspects of this general view on the nature of symbiotic 
stars were originally pointed by, for example, \cite{boy67}, 
\cite{tutyan76}, \cite{allen80}, \cite{stb}, \cite{kw84}, and 
\cite{nussvog87}. 

%
If the processes of mass loss, accretion, and ionization
are in mutual equilibrium, the symbiotic system releases
its energy at approximately a constant rate and spectral energy
distribution (SED). This stage is called
the {\em quiescent phase}. 
In most cases, the observed energy output of $\sim 10^3 - 10^4$\lo, 
is believed to be caused by stable hydrogen nuclear burning
on the WD surface, i.e. the accreted material burns as it is 
accreted \citep[e.g.][]{pacrud80,fuji82}. 
During quiescent phases, the main donor of particles into the 
circumbinary medium is represented by the wind from the giant. 
As a result, the symbiotic nebula is usually very extended, 
because the burning WD ionizes a large fraction of the giant's 
wind \citep[][]{stb,nussvog87}. Its origin allowed 
the mass loss rate from the giant to be estimated 
to a few $\times 10^{-7}$\myr\ 
\citep[e.g.][]{m+91,mio02,sk05}. Using the \ha\ method, 
\cite{sk06} found that the mass loss rate of the hot star 
during quiescent phases is significantly smaller, being of 
a few $\times 10^{-8}$\myr. 

Once the equilibrium between the primary processes is disturbed, 
e.g. by a transient increase in the accretion rate, the symbiotic 
system changes its radiation significantly, brightens 
up in the optical by typically 1--3 magnitudes, and shows 
signatures of enhanced mass outflow for a few months to years. 
This stage is called the {\em active phase}
\citep[e.g.][]{kenyon86}. However, a precise mechanism of 
the outbursts is not well understood to date. 
Recently, \cite{sok+06} suggested that nuclear shell burning 
on the WD could be enhanced by the influx of material 
from dwarf-nova-like disk instability, while \cite{bis+06} 
suggested a disruption of the disk as a result of variations 
in the wind velocity from the giant. 
The SED during active phases is characterized by a relatively 
cool, A--F stellar type of the continuum, a decrease 
(or disappearance) of highly ionized lines, and an increase 
of the nebular hydrogen emission. \cite{kenyon+91} and \cite{mk92} 
interpreted these features by an extended accretion disk 
around the hot star in CI~Cyg and AX~Per, whose plane is 
inclined to the observer with $\sim 73^{\circ}$. To avoid 
the problem of an unseen boundary layer, the authors speculated 
that the disk material expands out of the midplane and thus 
cools the boundary layer during the visual maximum. 
For the 1984-86 outburst of Z~And, \cite{fc+95} also indicated 
a cool UV continuum with a signature of the Rayleigh scattering 
attenuation around Ly-$\alpha$, along with a strong nebular continuum, 
whose emission measure increased by a factor of 3.4 with respect 
to quiescence. They interpreted these characteristics in terms 
of the ejection of an optically thick shell, which
substantially blocks the UV radiation. 
Finally, modelling the UV SED of active symbiotic binaries 
with a high orbital inclination confirmed the presence of such 
a two-temperature spectrum, where the cool component 
is produced by a relatively warm stellar source radiating at 
$\approx 20\,000$\,K, and the hot one is represented 
by a strong nebular radiation \citep[][]{sk05}. 
The former is not capable of producing the observed nebular 
emission, and thus the latter signals the presence of a hot 
ionizing source ($\ga 10^{5}$\,K) in the system, which is not 
seen directly by the observer. In addition, a strong 
Rayleigh scattering attenuation of the continuum around the 
Ly-$\alpha$ line directly indicates a significant number of 
neutral hydrogen atoms on the line of sight, which are not 
present during the quiescent phase. This effect was originally 
measured by \cite{fc+95}. 
Therefore, \citet{sk05} suggested that there is an edge-on 
disk around the accretor, the outer flared rim of which represents 
the warm pseudophotosphere. He also suggested that
the nebula is placed above/below the disk, since it is
ionized by the central hot star. 
A further important characteristic of the active phases is 
a significantly enhanced mass loss rate from the hot star 
\citep[e.g.][]{fc+95,nsv95,crok+02}.
Modelling the broad \ha\ 
wings, \cite{sk06} found that the mass loss rate of the active 
hot star enhances to a few times ($10^{-7} - 10^{-6}$)\myr, i.e. 
it exceeds the rate indicated during quiescent phases by a factor of 
$\ga 10$. The ejected material is ionized by the luminous 
central hot star, which thus enhances radiation from 
the symbiotic nebula \citep[see Tables 3 and 4 of][]{sk05}. 

These general properties of quiescent and active phases thus 
indicate a connection between the mass loss rate from the hot 
star and the level of the activity of symbiotic binaries. 
Recently, this connection was justified by \cite{sk+09}, who 
explained the anticorrelation between the supersoft X-ray and 
UV/optical fluxes due to the variable hot star wind during
different stages of activity of the symbiotic star AG~Dra. 

Accordingly, in this paper we investigate the idea that
the enhanced wind from the hot star can be responsible for 
the formation of the neutral disk-like zone during active 
phases. For this purpose, we applied the wind compression model, 
as introduced by \citet{bjorkcass93}, to explore how the wind 
from the WD can be compressed towards the equatorial plane 
due to its fast rotation. 
In Sect.~2, we outline in principle the wind compression 
model we use in our application, and calculate its ionization 
structure. Section~3 compares the model with observations and 
discusses the results. Conclusions are found in Sect.~4. 

\section{Ionization structure of the compressed wind}

Rotation of the hot star with radiation-driven winds leads to
compression of the outflowing material towards the equatorial regions.
The trajectory of each particle in the wind is determined by 
the gravity and radiation forces, which are both central forces. 
Due to conservation of the angular momentum vector, each particle 
launched at the surface of the star has to move in its own orbital 
plane, which is perpendicular to this vector. The mechanism can
be described by the wind compression model, in which the 
streamlines of the gas from both hemispheres of a rotating
line-driven wind are bent towards the equatorial plane 
\citep[e.g.][ and references therein]{lamcass99}. 
The flow of wind particles towards the equator enhances with 
rotation and weakens with radial distance from the star's surface. 
Two forms of equatorial density enhancements can be recognized: 
(i) If the flow from the hemispheres has a supersonic
component of the velocity perpendicular to the equatorial plane, 
a shock zone or disk will develop. This wind-compressed disk
(WCD) model was elaborated by \citet{bjorkcass93}. 
(ii) If the streamlines do not cross the equator (the so-called 
non-crossing trajectories) we get the wind-compressed zone
(WCZ) model described by \citet{igncassbjork96}. 
The shocks in the WCD model at the upper and lower 
boundary of the disk can produce very high temperatures 
\citep{bjorkcass93,krolray85}. As a result, the ionization 
structure in the wind would be influenced by the
presence of such a hot disk. 
In our approach, we restrict calculations for the sake
of simplicity to models with non-crossing trajectories, 
i.e. we apply the WCZ model. 

\subsection{Model assumptions}

The wind in our model begins at the surface of the hot WD 
with the radius $R_{\rm WD}$. It rotates with 
a velocity $v_{\rm rot}$, with the rotation axis perpendicular 
to the orbital plane. For this paper, 
we adopted $R_{\rm WD} = 0.01$\ro\ 
and $v_{\rm rot} = 100-600$\kms\ (see Sect.~2.3.1.).
The photons are emitted at a blackbody WD's pseudophotosphere 
with a temperature $T_{\star}$ and a radius $R_{\star} > R_{\rm WD}$. 
The pseudophotosphere represents the optically thick/thin 
boundary of the wind. It simulates the hot star in the 
model, which is characterized by the fundamental parameters $L_{\star}$, 
$R_{\star}$, and $T_{\star}$. Thus, the wind is accelerated deep 
inside the photosphere. As the wind mass loss rate varies 
between quiescent and active phases, $R_{\star}$ is also 
subject to variation. In the model, we assume that 
the WD's pseudophotosphere is spherical. 
%

From observations, we derive the fundamental parameters
$L_{\rm h}$, $R_{\rm h}^{\rm eff}$, $T_{\rm h}$ of the hot 
component in the binary (i.e. the object orbiting the 
cool giant). These parameters can deviate significantly from those 
of the hot star (i.e. $L_{\star}, R_{\star}, T_{\star}$ 
of the WD's pseudophotosphere as defined above), depending 
on the level of activity and the orbital inclination. 
For example, the  effective radius of the hot component, 
$R_{\rm h}^{\rm eff}$, is defined as the radius of a sphere, 
which has the same luminosity as the entire stellar-radiating 
disk-like object, derived by modelling the UV SED 
\cite[][ Eq.~(\ref{eqn:reff}) here]{sk05}. 
During active phases, $R_{\rm h}^{\rm eff}$ can be as large
as a few \ro\ for systems with a high orbital 
inclination, while during quiescent phases it is clustered 
around 0.1\ro\ only \citep[e.g.][]{sk05,sk+11}. When viewing 
the system more from the pole, the effective radius shrinks 
significantly. An example is AG~Dra, 
for which the mean values of 
$R_{\rm h}^{\rm eff} = 0.063 \pm 0.013$\ro\ and 
                      $0.033 \pm 0.010$\ro\ 
were derived by modelling the SED during active and quiescent 
phases, respectively \citep[][]{sk05,sk+09}. 
This suggests that we measure $R_{\rm h}^{\rm eff} \gg R_{\star}$
for eclipsing systems and $R_{\rm h}^{\rm eff} \approx R_{\star}$
for non-eclipsing systems. 
Accordingly, in the model calculations we adopted 
$R_{\star} = 0.06$\ro\ for active phases. Other parameters of 
the hot star and the wind that are needed for the model
calculations are introduced in Sect.~3.1. 

Finally, we neglect the effect of the orbital motion on the 
shaping of the compressed wind formation. For a typical orbital 
period of $\sim 2$ years and total mass of $\sim 2$\,\mo\ of 
a symbiotic binary, the orbital velocity of its components
is only $v_{\rm orb} \sim 30$\kms. As $v_{\rm orb}$ is
significantly smaller than the terminal velocity of the wind
from the WD $v_\infty$ (see Sect~2.3.2), 
the momentum flux $\dot M (v_\infty + v_{\rm orb}) \approx 
\dot M (v_\infty - v_{\rm orb})$ for a constant $\dot M$ from 
the WD. Therefore, the density gradient at the front and back 
side of the WD in the direction of its orbital motion will 
also be comparable, and the expected prolongation of the 
compressed wind formation will probably not be significant. 
As a result, we ignore the orbital motion of the binary and 
calculate the ionization structure in the compressed wind for 
a stationary situation and pure hydrogen gas, similar to the 
\cite{stb} calculation of the ionization structure 
for the quiescent phase.

\subsection{Density in the WCZ model}

According to \cite{bjorkcass93}, in a star-centred spherical 
coordinate system ($r, \theta, \phi$), the wind compression model 
assumes an azimuthal symmetry, i.e. the density of the wind does 
not depend on the azimuthal angle $\phi$. Therefore, the density 
is a function of the radial distance $r$ from the star's centre 
and the polar angle $\theta$, measured from the spin axis. 
The wind density distribution follows from the mass continuity 
equation as 
\begin{equation}
  N_{\rm H}(r,\theta) =
          \frac{\dot{M}}{4\pi r^2 \mu_{\rm m} m_{\rm H}
          v_{\rm r}(r)}\left(\frac{d\mu}{d\mu_{0}}\right)^{-1},
\label{eqn:nh}
\end{equation}
where $\dot{M}$ is the mass loss rate of the star, $\mu_{\rm m}$ 
is the mean molecular weight, $m_{\rm H}$ is the mass of 
the hydrogen atom, $v_{\rm r}(r)$ is the radial component of 
the wind velocity. We used the $\beta$-law wind as introduced 
by \cite{lamcass99}, 
\begin{equation}
  v_{\rm r}(r) = v_{\infty}\left(1-\frac{bR_{\rm {WD}}}{r}
                           \right)^{\beta},
\label{eqn:betalaw}
\end{equation}
where $R_{\rm {WD}}$ is the origin of the wind, $\beta$
characterises an acceleration of the wind (i.e. how steep 
the velocity law is), and the parameter $b$ is given by 
\begin{equation}
  b = 1-\left(\frac{a}{v_{\infty}}\right)^{1/\beta},
\label{eqn:a}
\end{equation}
where $a$ is the initial velocity of the wind at its origin 
(see Sect.~2.3.2). 
In comparison with \cite{bjorkcass93}, we assume that 
$v_{\infty}$ does not depend on the polar angle $\theta$. 
The geometrical factor $d\mu/d\mu_{0}$ describes the 
compression of the wind due to rotation of the star 
\citep[see][]{bjorkcass93,igncassbjork96}. 
A summary of the model can also be found in the textbook 
of \cite{lamcass99}. 

Knowing the density distribution in the compressed wind 
(Eq.~\eqref{eqn:nh}), we can determine the ionization structure 
using the equation of photoionization equilibrium. 

\subsection{Ionization boundaries in the WCZ model}

In this section, we calculate the ionization boundaries in
the wind from the hot star. The ionization boundary is defined
by the locus of points at which ionizing photons are completely
consumed along the path outward from the ionizing star. For 
the sake of simplicity, we restrict our calculations to 
a wind containing only hydrogen atoms and assume Case B
recombination \citep[e.g.][]{ost74}. 

At each point of the nebula, the photoionization equilibrium
equation can be written as
\begin{equation}
  N_{\rm H^{0}}(r,\theta) \int_{\nu_{0}}^{\infty}
  \frac{4\pi J_{\nu}}{h\nu}a_{\nu}{\rm d}\nu =   
           N_{\rm p}(r,\theta) N_{\rm e}(r,\theta)
           \alpha_{\rm B}(T_{\rm e}),
\label{eqn:equi1}
\end{equation}   
where $J_{\nu}$ is the mean intensity of radiation,
$N_{\rm H^{0}}(r,\theta)$, $N_{\rm p}(r,\theta)$ and
$N_{\rm e}(r,\theta)$ are neutral hydrogen, proton and electron
densities by number per unit volume, respectively, $a_{\nu}$ is
the ionization cross section for hydrogen, $h$ is the Planck's 
constant, $\nu_{0}$ is the ionization frequency for hydrogen   
atoms, and $\alpha_{\rm B}(T_{\rm e})$ is the total hydrogenic
recombination coefficient in case B. 

The left-hand side of Eq.~(\ref{eqn:equi1}) gives the number 
of ionizations and the right-hand one gives the number of 
recombinations per unit volume per unit time. Their equivalence 
expresses the condition for the ionization equilibrium. 
If the absorption and geometrical dilution are taken into account,
the mean intensity of the stellar radiation can be written as 
\begin{equation}
  4\pi J_{\nu} = 4\pi W(r)B_{\nu}(T_{\star})e^{-\tau_{\nu}},
\end{equation}
where $\pi B_{\nu}(T_{\star})$ is the flux at the surface of
the hot star (i.e. the WD's pseudophotosphere, see Sect.~2.1.), 
$\tau_{\nu}$ is the radial optical depth for hydrogen, and 
$W(r)$ is the geometrical dilution coefficient, which represents 
the ratio of the solid angle, subtended by the central star at 
the point of observation, to $4\pi$ \citep[e.g.][]{gur97} 
\begin {equation}
  W(r) = \frac{1}{2}\left[1-\sqrt{1-\left(\frac{R_{\star}}{r}
         \right)^{2}}\right],
\end {equation}
where $R_{\star}$ is the radius of the source of the ionizing 
photons. Since the luminosity of the hot star per unit frequency 
interval is given by 
\begin{equation}
  L_{\nu} = 4\pi R_{\star}^{2} \pi B_{\nu}(T_{\star}),
\end{equation}
we can rewrite Eq.~(\ref {eqn:equi1}) into the form
\begin{equation}
  N_{\rm H^{0}}(r,\theta) \int_{\nu_{0}}^{\infty} \frac{L_{\nu}}{h\nu}
  \frac{W}{\pi R_{\star}^{2}} a_{\nu} e^{-\tau_{\nu}} {\rm d}\nu
  = N_{\rm p}(r,\theta) N_{\rm e}(r,\theta) \alpha_{\rm B}(T_{\rm e}).
\label{eqn:equi2}  
\end{equation}     
Denoting the rate of photons from the hot star
capable of ionizing hydrogen as 
\begin{equation}
  L_{\rm H} = \int_{\nu_{0}}^{\infty} \frac{L_{\nu}}{h\nu}{\rm d}\nu
            = 4\pi R_{\star}^{2} \int_{\nu_{0}}^{\infty}
              \frac{\pi B_{\nu}(T_{\star})}{h\nu} {\rm d}\nu
\label{eqn:LH} 
\end{equation} 
and integrating both sides of Eq.~(\ref {eqn:equi2}) over
the radial distance $r$, we get 
\begin{equation}
  \frac{L_{\rm H}}{2\pi R_{\star}^{2}} =
  \int_{R_{\star}}^{r_{\theta}} 
  \frac{N_{\rm p}(r,\theta) N_{\rm e}(r,\theta)
                            \alpha_{\rm B}(T_{\rm e})}
       {1-\sqrt{1-\left(\frac{R_{\star}}{r}\right)^{2}}} {\rm d}r,
\label{eqn:equi3} 
\end{equation}    
where $r_{\theta}$ is the radius of the \ion{H}{i}/\ion{H}{ii}
boundary at the direction $\theta$. To obtain the whole 
ionization boundary, one needs to solve 
Eq. (\ref{eqn:equi3}) for each direction, given by the polar
angle $\theta$.
Due to the symmetry of the wind with respect to the equatorial
plane and considering just the non-crossing streamlines of    
the gas, it is sufficient to perform calculations from        
the star's pole ($\theta=0$) to its equator ($\theta=\pi/2$). 
As in the classical Str\"{o}mgren sphere calculations, we
assume that all atoms of hydrogen with density distribution
$N_{\rm H}(r,\theta)$ are ionized at the ionization radius
$r_{\theta}$ (i.e. $N_{\rm p} = N_{\rm e} \approx N_{\rm H}$) 
and neutral outside it (i.e. $N_{\rm p} = N_{\rm e} \approx 0$).
That means that within the ionization boundary we can take 
$N_{\rm p}N_{\rm e} = N_{\rm H}^{2}$. 
\citet{nussvog87} expressed
the ionization equilibrium equation in the wind with spherical
density distribution in the form 
\begin{equation}
  \frac{L_{\rm H}}{4\pi} =
  \int_{R_{\star}}^{r_{\theta}} N_{\rm p}(r) N_{\rm e}(r)
  \alpha_{\rm B} (T_{\rm e}) r^{2} {\rm d}r,
\end{equation}
which can be derived from Eq.~(\ref{eqn:equi1}) if we adopt
the point source approximation (i.e. the dilution factor   
$W=1/4(R_{\star}/r)^{2}$). This approximation is valid for 
$R_{\star}/r \ll 1$. 
However, in our treatment, the ionization boundaries can be
close to the central star (especially in the equatorial plane),
which requires using Eq.~(\ref{eqn:equi3}) to calculate 
correctly the \ion{H}{i}/\ion{H}{ii} boundaries, $r_{\theta}$.

Finally, we express the radial distance $r$ in units of
$R_{\rm {WD}}$, and define $u$ as 
\begin{equation}
  u \equiv \frac{r}{R_{\rm {WD}}}.
\label{eqn:par}
\end{equation} 
Then, with the further approximation of assuming a constant
$T_{\rm e}$ and thus a constant $\alpha_{\rm B}$ throughout
the ionized nebula, Eq.~(\ref{eqn:equi3}) with the aid of  
Eq.~(\ref{eqn:nh}) can be expressed as
\begin{equation}
  X = \int_{\frac{R_{\star}}{R_{\rm{WD}}}}
          ^{u_{\theta}} \frac{{\rm d}u}{u^4
      \left(1-\frac{b}{u}\right)^{2\beta}  
      \left(\frac{d\mu}{d\mu_0}\right)^2   
      \left(1-\sqrt{1-\left(\frac{R_{\star}}{R_{\rm{WD}}}
          \frac{1}{u}\right)^2} \right)},
\label{eqn:calcX} 
\end{equation}    
where $u_{\theta} = r_{\theta}/R_{\rm {WD}}$ and the parameter
$X$ is given by
\begin{equation}
  X = \frac{8\pi \mu_{\rm m}^{2} m_{\rm H}^{2}}{\alpha_{\rm B}(T_{\rm e})}
  \frac {R_{\rm {WD}}^{3}}{R_{\star}^{2}} L_{\rm H}
  \left(\frac{v_{\infty}}{\dot{M}}\right)^{2}. 
\label{eqn:X}  
\end{equation} 
The radius of the hot star $R_{\star}$ is determined in terms
of its luminosity $L_{\star}$ and temperature $T_{\star}$ as 
\begin{equation}
  R_{\star} = \sqrt {\frac{L_{\star}}{4\pi \sigma T_{\star}^{4}}},
\label{eqn:R}
\end{equation}
where $\sigma$ is the Stefan-Boltzmann constant. 
Solutions of Eq.~(\ref{eqn:calcX}) for $u_{\theta}$ at 
directions $\theta$ define the \ion{H}{i}/\ion{H}{ii} 
boundary. First, we introduce some critical parameters 
of the rotating WD and its wind. 

\subsubsection{Rotational velocity of the WD}

Rotational velocities of the accretors in symbiotic binaries 
are not commonly known. 
It was observationally found that isolated WDs rotate with 
$v_{\rm rot}\sin i \la$ 40\kms\ \citep[e.g.][]{heber97,koester98}. 
Therefore, it is reasonable to assume that accreting WDs 
will rotate faster than isolated WDs because of the angular 
momentum transfer from the accreting material. 

Direct evidence for a rapid rotation of accreting WDs is 
complicated due to a high-density circumstellar medium that 
shades photospheric absorption lines, which otherwise might be 
used to derive a possible rotation speed. An exception 
is the case of the WD in the dwarf nova VW~Hyi, for which 
\cite{sion+95} derived a rotational velocity of 
$v_{\rm rot}\sin i \simeq 600$\kms
by fitting the broad, shallow \ion{Si}{iv} absorption doublet 
in its high-resolution Hubble Space Telescope spectrum. 
In cases where the energy of the accretor comes solely from 
the accretion process, the fraction produced by the boundary layer 
depends on the angular velocity of the accretor. It was found 
that the observed emission from a boundary layer in cataclysmic 
variables is significantly lower (around one-quarter) than 
the flux emitted by the accretion disk, which can be explained 
by a rapid rotation of the WD accretor 
\citep[e.g.][]{hoare+drew91,belloni+91,vrtilek+94}. 
Considering the spinning up the accreting star, 
\cite{pop+nar95} derived a theoretical expression for the 
boundary layer luminosity. 
Comparing the boundary layer luminosity, inferred from 
observations, with the theoretical one, \cite{sk+05} found 
that the WD in the symbiotic star EG~And rotates at 0.3--0.5 
of its critical velocity. 
Further support for the rapid rotation of accreting WDs is 
provided by coherent oscillations in the $U$-light curve of 
the symbiotic star CH~Cyg with the period of 330 -- 500\,s, 
observed independently during three nights of its quiescent phase 
\citep[][]{mikolajewski+90}. The authors interpreted this 
periodic variability as due to rotation of a magnetic WD 
in the system. 

Accordingly, we first investigated cases with very fast rotational 
velocities of a WD. However, we found that 
$v_{\rm rot} \gtrapprox 600$\kms\ leads to a high compression 
of the wind, whose particles can cross the equatorial plane 
for appropriate parameters of the WD wind. This, however, does not 
satisfy assumptions of the WCZ model. Therefore, to calculate 
the ionization boundaries within the WCZ model, we examined 
cases for $v_{\rm rot} \la 600$\kms. 
%
\begin{figure*}
\begin{center}
\resizebox{\hsize}{!}{\includegraphics[angle=-90,trim=0.3cm 3.5cm 0.3cm
4cm]{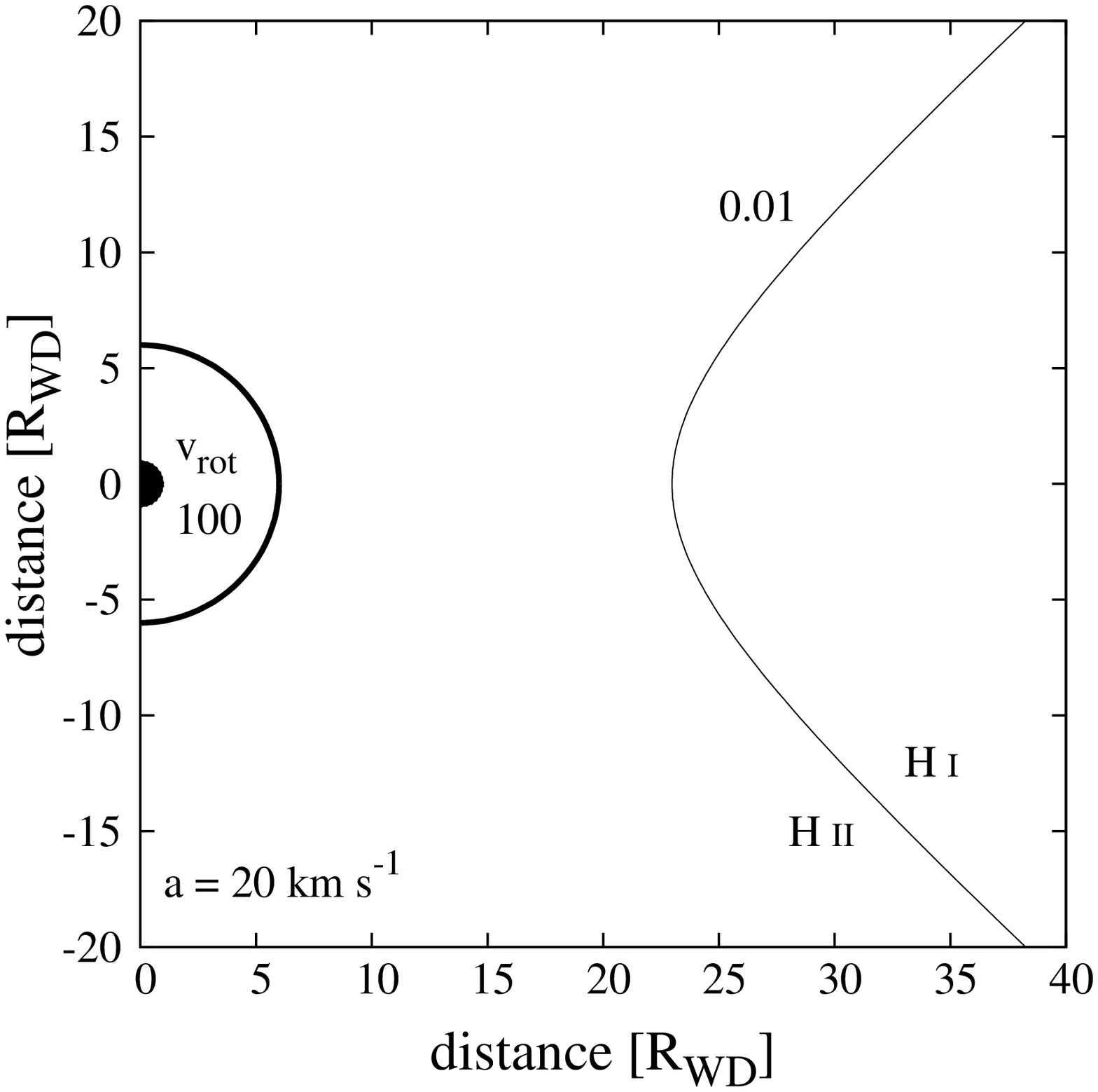}
                      \includegraphics[angle=-90,trim=0.3cm 3.5cm 0.3cm
4cm]{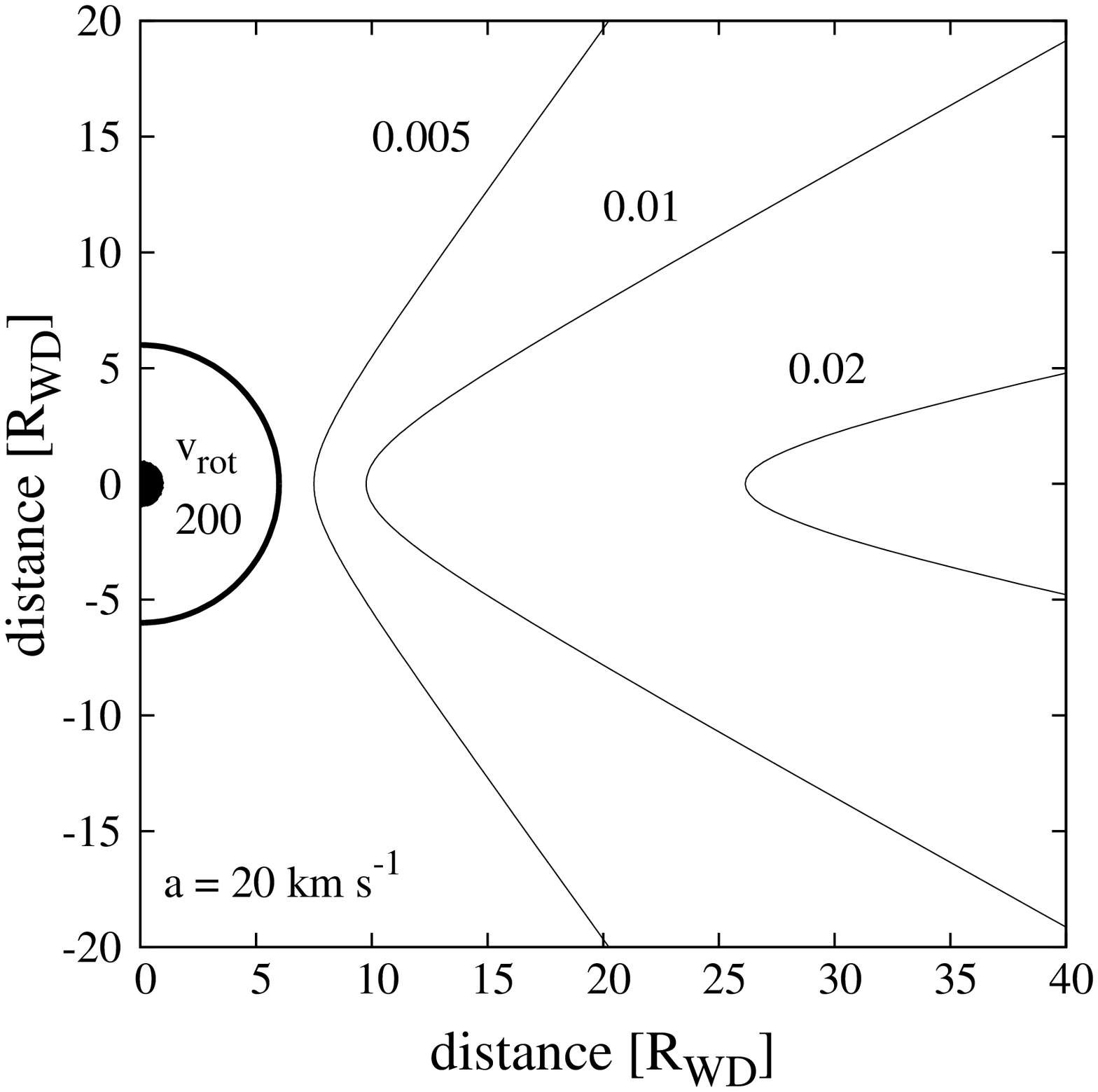}
                      \includegraphics[angle=-90,trim=0.3cm 3.5cm 0.3cm
4cm]{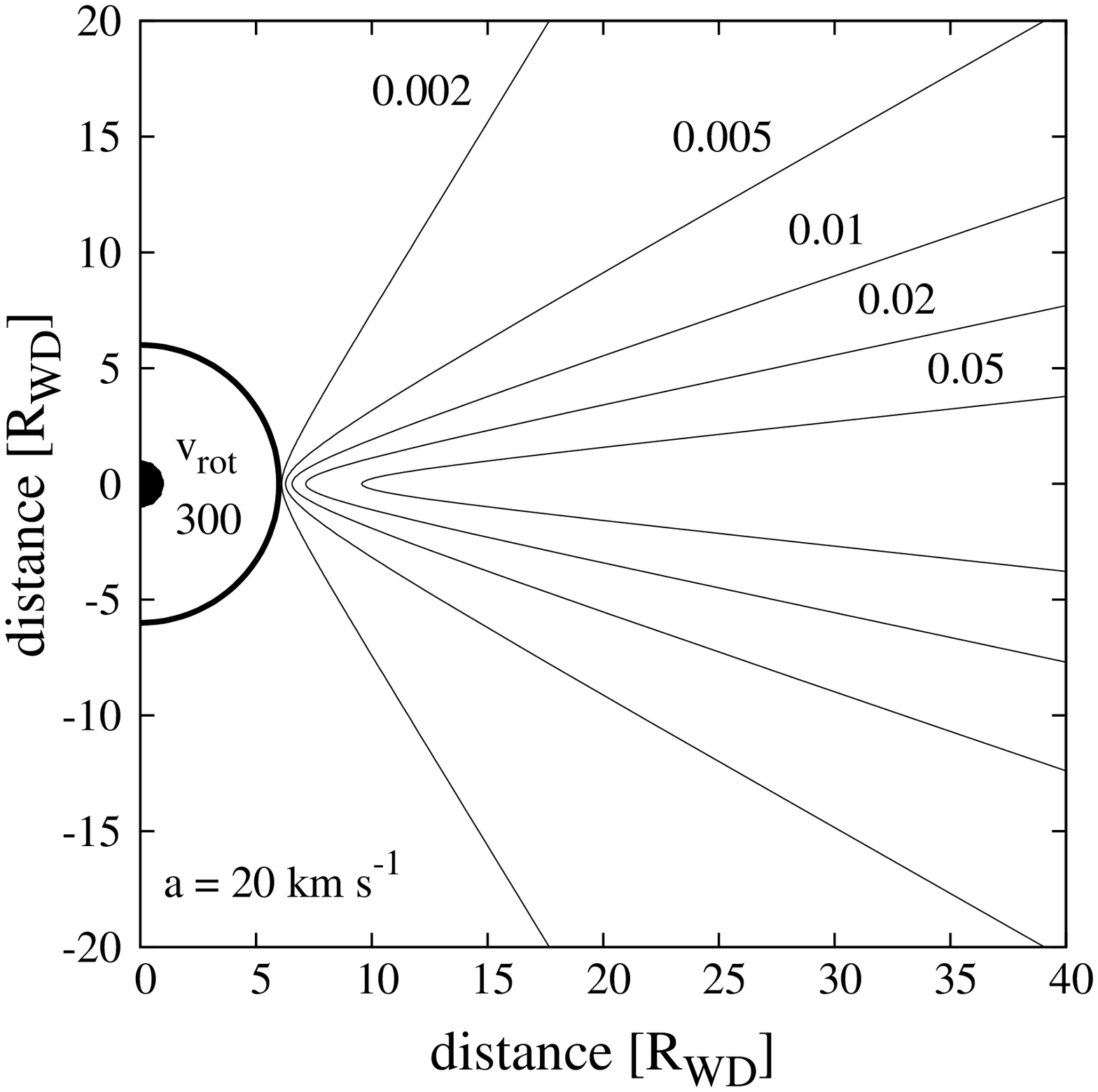}}
\resizebox{\hsize}{!}{\includegraphics[angle=-90,trim=0.3cm 3.5cm 0.3cm
4cm]{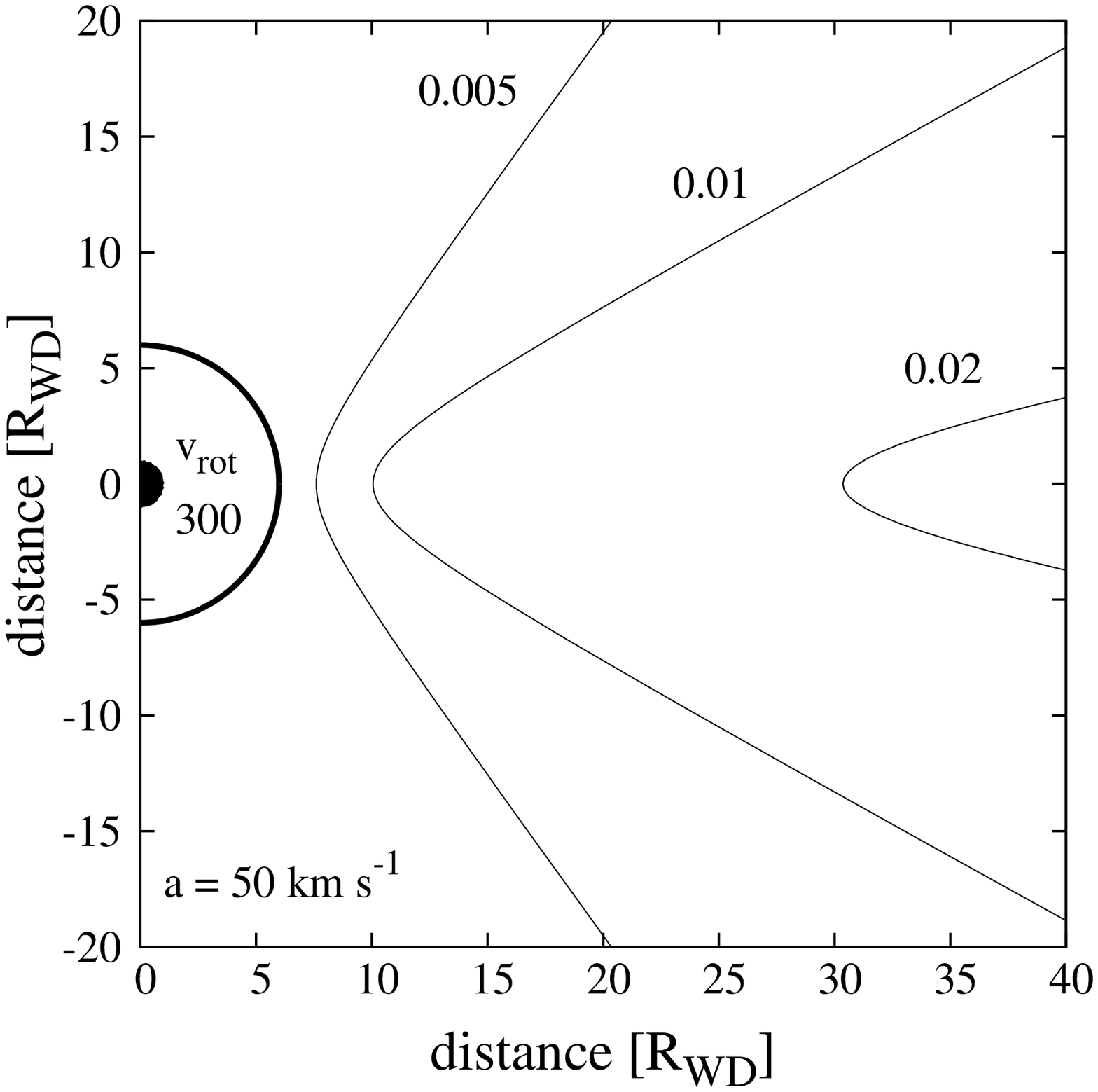}
                      \includegraphics[angle=-90,trim=0.3cm 3.5cm 0.3cm
4cm]{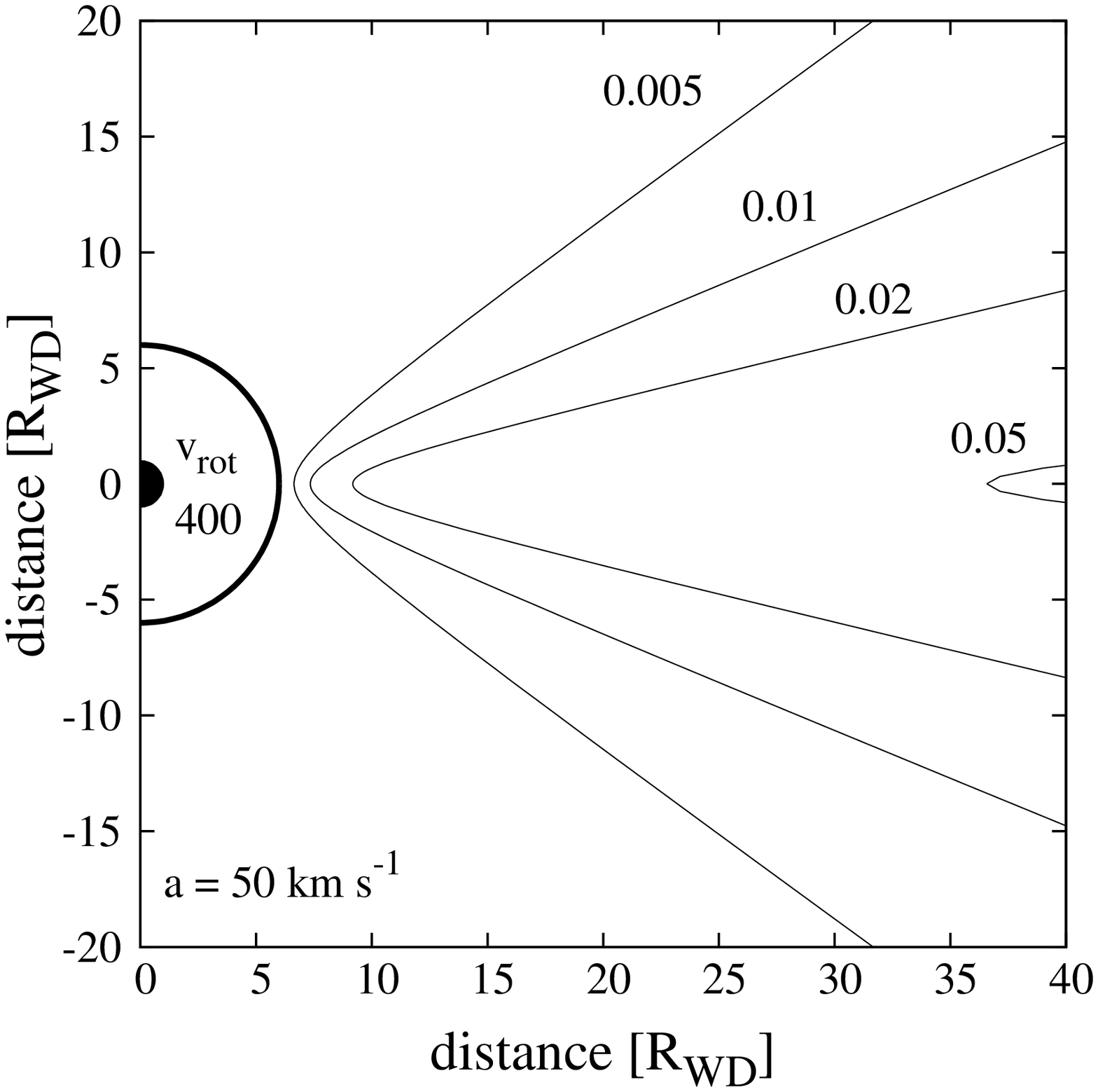}
                      \includegraphics[angle=-90,trim=0.3cm 3.5cm 0.3cm
4cm]{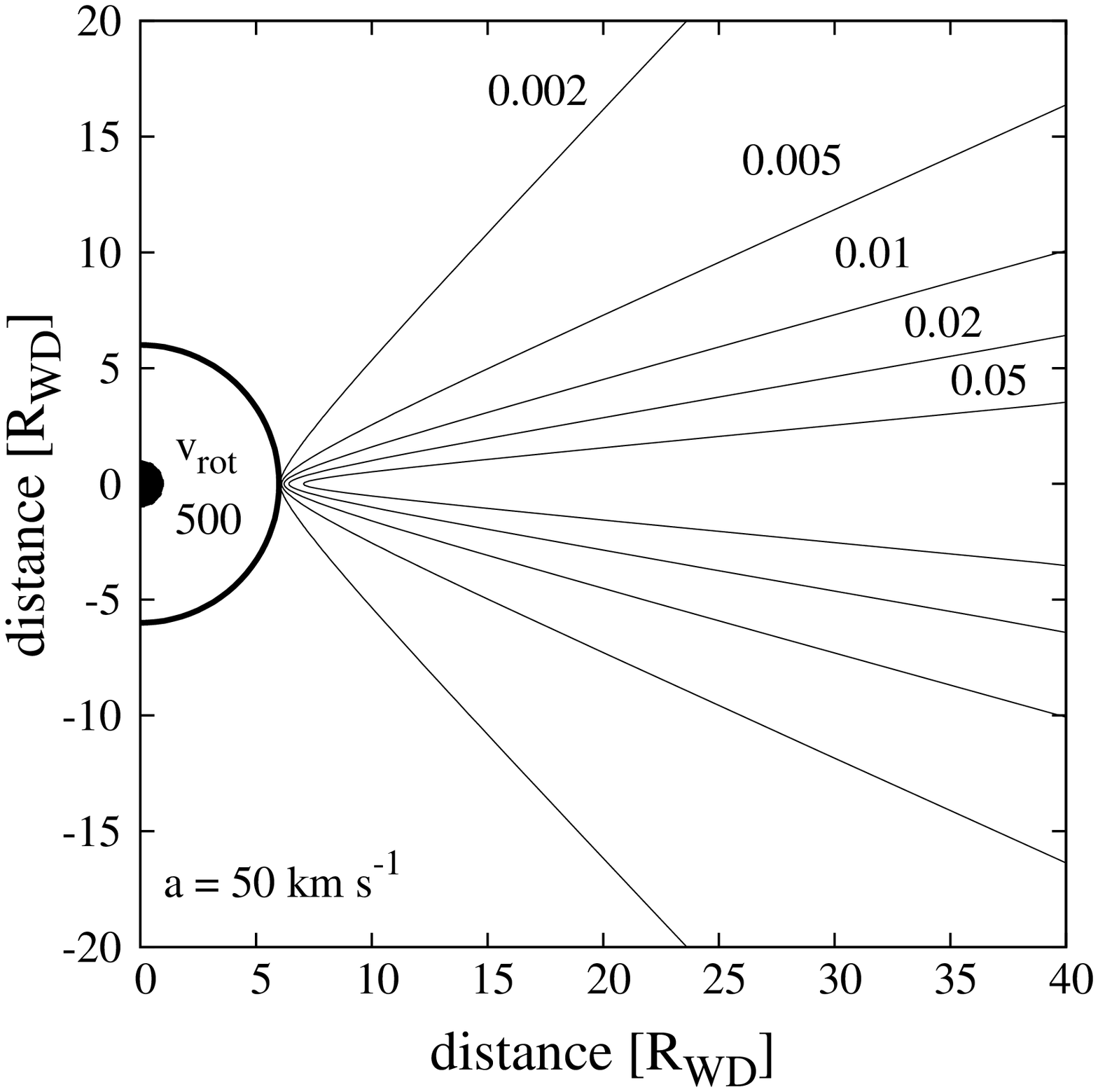}}
\resizebox{\hsize}{!}{\includegraphics[angle=-90,trim=0.3cm 3.5cm 0.3cm
4cm]{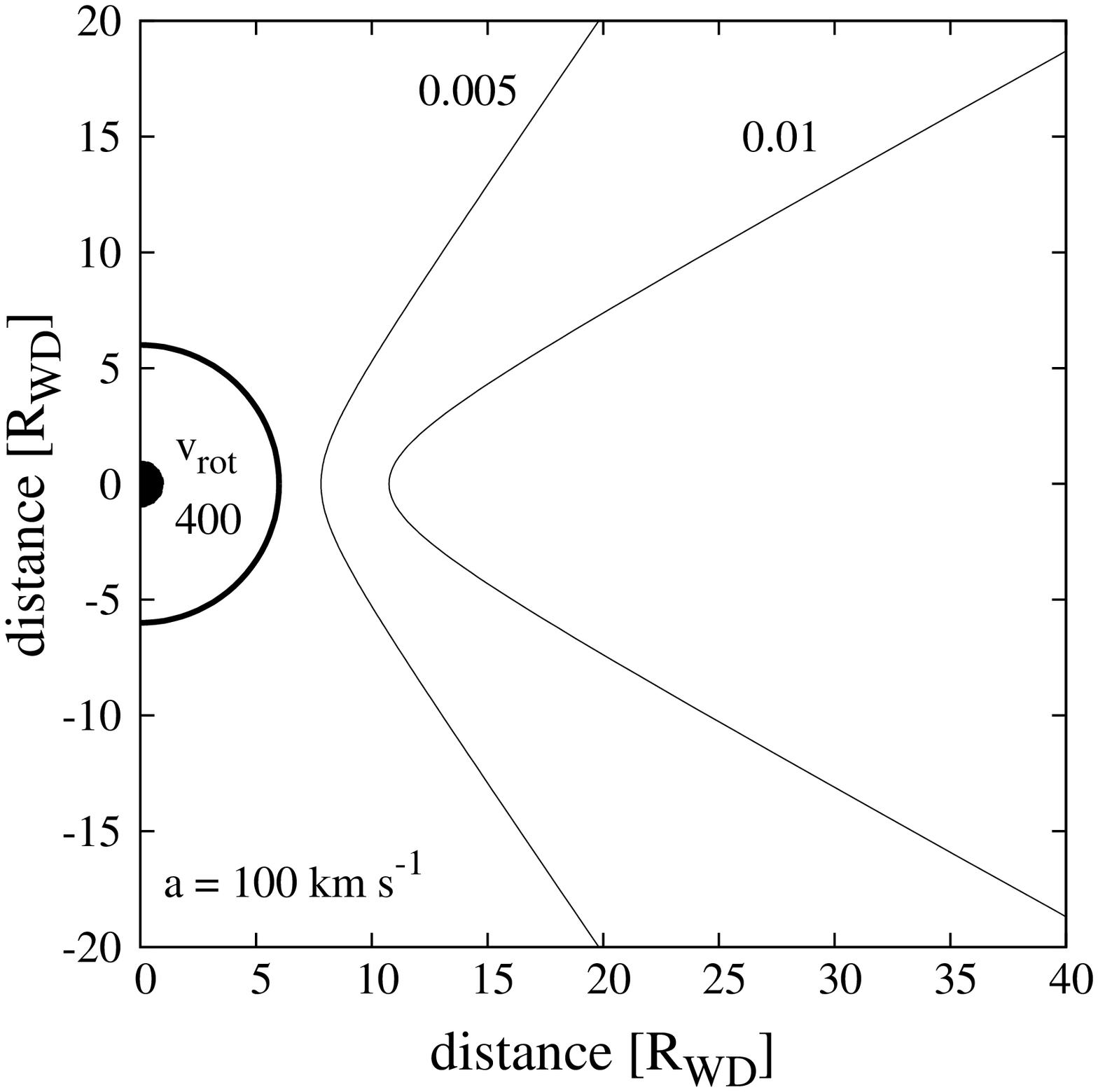}
                      \includegraphics[angle=-90,trim=0.3cm 3.5cm 0.3cm
4cm]{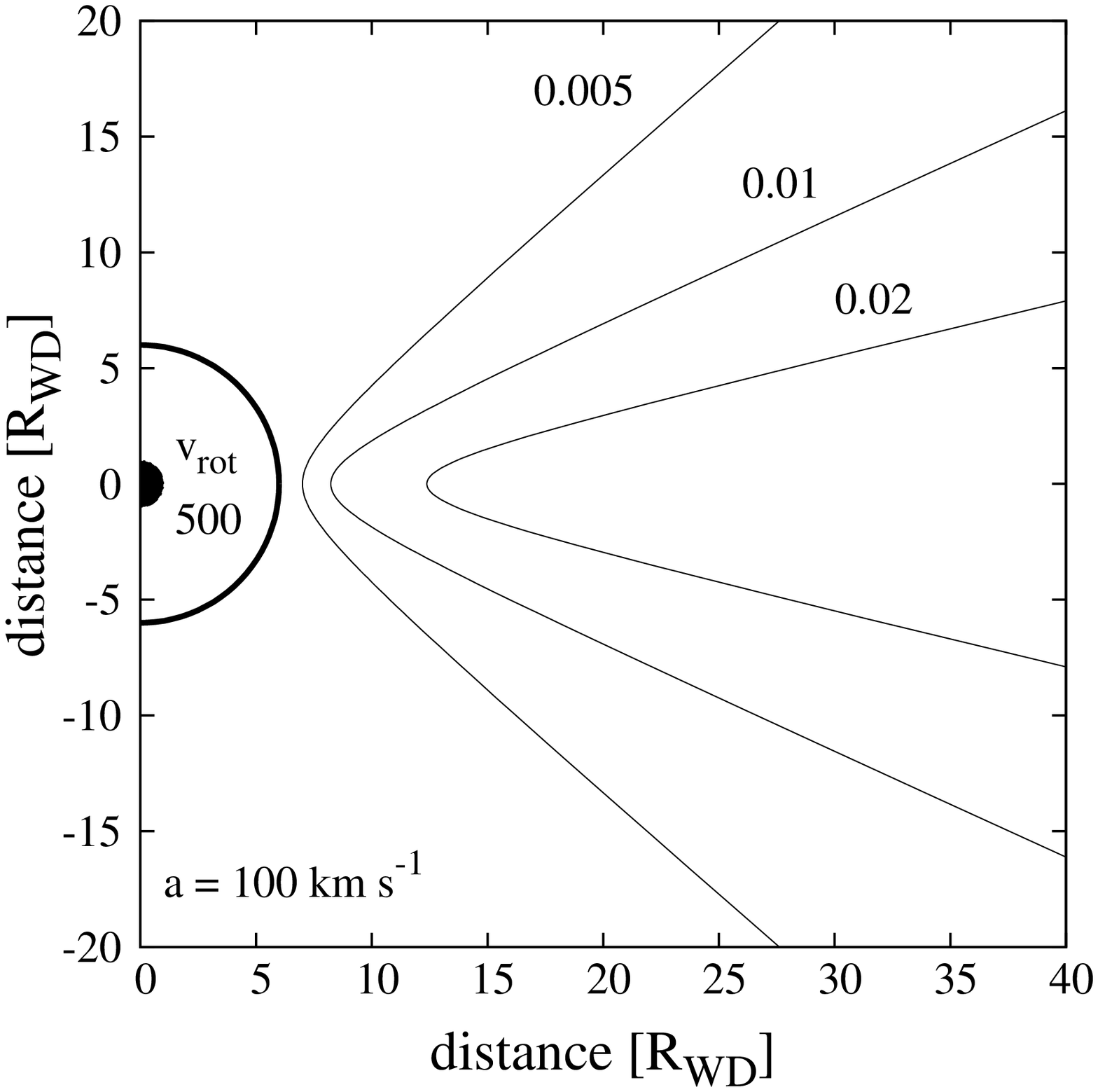}
                      \includegraphics[angle=-90,trim=0.3cm 3.5cm 0.3cm
4cm]{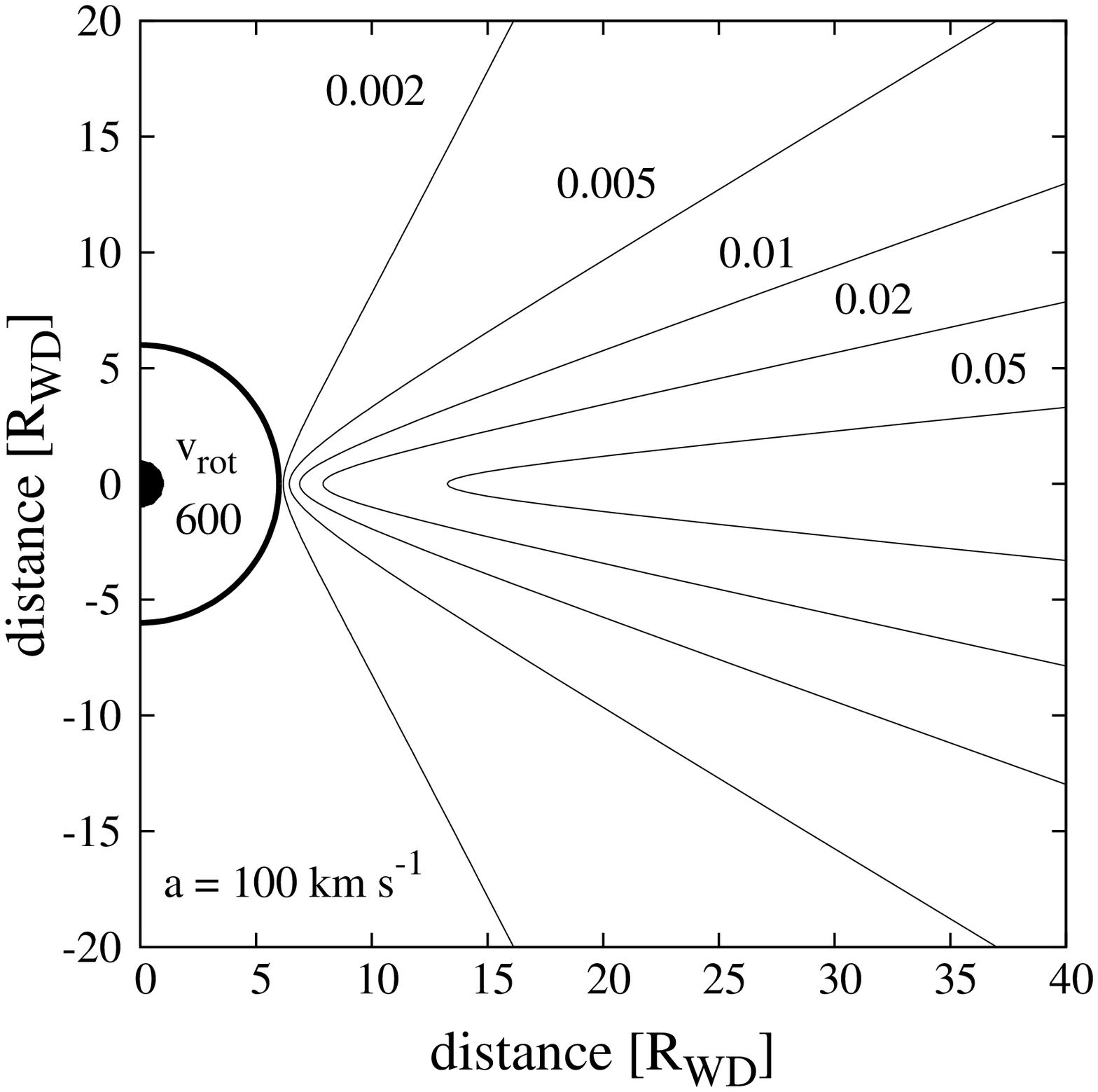}}
\end{center}
\caption[]{
\ion{H}{i}/\ion{H}{ii} boundaries in the wind compression 
model (Sect.~2) at the vicinity of the hot star (solid thick 
circle). They were calculated according to 
Eq.~(\ref{eqn:calcX}) for $a$ = 20, 50, and 100\kms, displayed 
at the top, middle, and bottom row of panels, respectively,
with fixed $v_{\infty} = 2\,000$\kms\ and $\beta =1.7$. 
The rotational velocities, $v_{\rm rot}$, are in \kms. 
The boundaries are labelled by the corresponding parameter $X$. 
The neutral \ion{H}{i} zone extends from the boundary in 
directions outside the star. 
The remainder part of the wind is ionized. Boundaries are 
axially symmetric with respect to the spin axis of the WD 
(the black circle), assumed to be perpendicular to the orbital 
plane. Distances are in units of the WD radius. 
          }
\label{fig:h1h2}
\end{figure*}

\subsubsection{Parameters of the hot star wind}

Modelling the broad \ha\ wings observed in symbiotic binaries 
by an optically thin bipolar wind from their hot components, 
\cite{sk06} derived mass loss rates of a few 
$\times 10^{-8}$\myr\ and of a few 
$\times\,(10^{-7} - 10^{-6})$\myr\ during quiescent and 
active phases, respectively. The acceleration parameter 
$\beta = 1.7 - 1.8$ was in effect for both stages, 
while the wind terminal velocity during quiescence, 
$v_{\infty}\sim 1\,000$\kms increased to $\sim 2\,000$\kms, 
during activity. 

For the initial velocity $a$ of the wind in the model (Eq.~(3)), 
we adopted the speed of sound at the interior 
parts of the ionized wind. According to modelling the extensive 
wings of emission lines from highly ionized elements by electron 
scattering, the electron temperature at the vicinity of the hot 
star, $T_{\rm e} = 30\,000 - 40\,000$\,K 
\citep[see Fig.~1 of][]{seksko12}, corresponds to the speed of 
sound $a \sim 22$\kms\ (monoatomic gas from H and 
the adiabatic index of 5/3 were assumed). 
This value probably represents a lower limit, because $T_{\rm e}$ 
within the acceleration zone, i.e. in the vicinity of the burning 
WD during outbursts ($T_{\star} \ga 10^5$\,K), can be considerably 
higher. Furthermore, because of the increase of the hot star 
luminosity, the terminal velocity of the wind moves from 
$v_{\infty} \sim 1\,000$ to $\sim 2\,000$\kms\ during active 
phases. Thus one can also expect an equivalent increase in 
the initial velocity of the radiatively driven wind.
Accordingly, we adopted $a$ = 20, 50 and 100\kms\ to calculate 
the \ion{H}{i}/\ion{H}{ii} boundaries in the wind within the 
WCZ model. 

\subsection{Results}

We calculated the \ion{H}{i}/\ion{H}{ii} boundaries, $u_{\theta}$, 
according to Eq.~(\ref{eqn:calcX}) for $R_{\star}/R_{\rm WD} = 6$
(Sect.~2.1.), $v_{\rm rot} \la 600$\kms\ (Sect.~2.3.1.), 
$v_{\infty} = 2\,000$\kms, $\beta = 1.7$, and $a = 20, 50, 100$\kms\ 
(Sect.~2.3.2.) with the aim of finding a range of the parameter $X$, 
for which a neutral disk-like zone can be created. The results 
are shown in Fig.~\ref{fig:h1h2}, which demonstrates that for 
\begin{equation}
          0.002 \lessapprox X \lessapprox 0.05
\label{eqn:xrange}		  
\end{equation}
a neutral disk-like zone can be created at the equator of the 
hot star, flaring away from it. For 
$X \lessapprox 0.002$, the ionization boundary tends to enclose 
the star, while $X \gtrapprox 0.002$ moves the boundary to 
farther distances with its simultaneous narrowing, until it 
disappears for $X\gtrapprox 0.05$. 
For a given set of parameters, different compression of the 
outflowing material is caused by a different WD rotation. 
Higher rotational velocities lead to a higher compression of 
the wind towards the equatorial plane and to wider range of 
the parameter $X$, resulting in creation of the neutral 
disk-like zone (see Fig.~\ref{fig:h1h2}). 

Shaping of the ionization boundaries is also sensitive to the 
wind parameters $a$, $v_{\infty}$, and $\beta$. For a higher 
initial velocity $a$ or terminal velocity $v_{\infty}$, the 
corresponding ionization boundary lies farther from the hot star.
This is because higher velocities lead to lower densities in the wind, 
which moves the point of the balance between the flux of ionizing 
photons and neutral atoms to farther distances. 
On the other hand, a higher value of the parameter $\beta$ 
accelerates the wind from $a$ to $v_{\infty}$ more slowly, 
leading to higher densities in the wind and thus to a closer 
position of the neutral zone to the hot star, which is 
characterized by a larger opening angle.

The range of the ionization parameter $X$, for which the neutral 
disk-like zone can be created, is important for comparison 
with observations. 

\section{Comparison with observations}

Application of the wind compression model to the hot components 
of symbiotic binaries during active phases revealed the 
possibility of forming a neutral disk-like zone around the hot 
star. This confirmed theoretically the previous suggestion 
based on modelling the UV/near-IR SED that there is a neutral 
flared disk surrounding the active hot star at the orbital 
plane \citep[see Fig.~27 of][]{sk05}. 
The model is also consistent with the ionization structure 
derived from the photometric and spectroscopic analysis of 
the eclipsing symbiotic binary AX~Per during its recent active 
phase \citep[see Fig.~9 of][]{sk+11}. 

In the following sections, we verify the model by comparing its 
basic properties with those obtained independently from 
observations. 
In Sect.~3.1., we investigate the dependence of the wind 
compression, given by the parameter $X$, on the mass loss rate 
from the hot star. The aim is to determine if the presence or 
absence of the neutral disk around the hot stars during active 
or quiescent phases is consistent with the model predictions. 
In Sect.~3.2., we compare the model column density of the neutral 
hydrogen, calculated throughout the neutral zone, with that 
derived from observations. Finally, in Sect.~3.3., we compare 
the observed emission measure to that given by the model.

\subsection{Wind compression as a function of $\dot M$}

According to Eq.~(\ref{eqn:calcX}), the ionization structure 
in WCZ model is given by the parameter $X$. It reflects 
a degree of compression of the wind (Fig.~\ref{fig:h1h2}). 
The changes of the $X$ parameter are dominated by significant 
changes in the mass loss rate from the hot star during 
different phases of the star's activity (Sect.~2.3.2.). 
Therefore, to probe if the presence or absence of the neutral 
disk in a symbiotic system is consistent with the WCZ model, we 
investigate a dependence of $X$ on $\dot M$ for physical 
parameters of selected symbiotic stars during quiescent and 
active phases. According to Eq.~(\ref{eqn:X}), the value of 
the parameter $X$ is given by the hot star parameters
$L_{\rm H}$, $R_{\star}$ and those of its wind, $\dot M$, 
$v_{\infty}$. 
%
\begin{figure}
\begin{center}
\resizebox{\hsize}{!}{\includegraphics[angle=270]{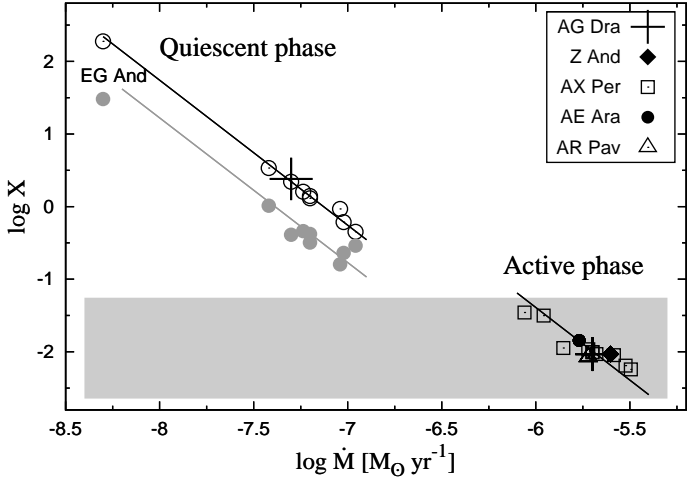}}
\end{center}
\caption[]{
Parameter $X$ as a function of $\dot{M}$ from the hot star, 
as given by physical parameters of the selected symbiotic 
stars during quiescent (left) and active (right, keys) 
phases. Their fits with Eq.~(\ref{eqn:xmd}) are denoted by 
the solid lines. The grey belt corresponds to 
$X = 0.002 - 0.05$, which leads to creation of the neutral 
zone around the hot star in the form of a flared disk 
(see Fig.~\ref{fig:h1h2}, Sects.~2.3. and 2.4.). 
         }
\label{fig:mdot}
\end{figure}

\subsubsection{$X(\dot M)$ relationship during quiescent phases} 

For a quiescent phase, we selected ten objects with available 
$L_{\rm h}$, $R_{\rm h}^{\rm eff}$, $T_{\rm h}$ of their hot 
components, as derived by modelling the UV/near-IR SEDs 
\citep[EG~And, Z~And, CI~Cyg, LT~Del, YY~Her, RW~Hya, SY~Mus, 
AG~Peg and AX~Per; see Table~3 of][]{sk05} and X-ray/near-IR SED 
\citep[AG~Dra; see Table~3 of][]{sk+09}. 
$L_{\rm H}$ is determined by $L_{\rm h}$ and $T_{\rm h}$. 
Parameters of the wind were taken from Table~1 of \cite{sk06}. 
For AG~Dra and LT~Del, we adopted average values
  $v_{\infty} \equiv 1000$\kms\ and 
  $\dot M \equiv 5\times 10^{-8}$\myr. In the case of EG~And, 
we used 
  $v_{\infty} = 900$\kms\ and 
  $\dot M = 5\times 10^{-9}$\myr, according to \cite{murset+97}. 

To get a rough estimate of the $X(\dot M)$ quantities, we first 
used fundamental parameters $L_{\rm h}$, $R_{\rm h}^{\rm eff}$, 
$T_{\rm h}$ as derived from observations. The results are plotted 
in Fig.~\ref{fig:mdot} by full grey circles. The `plus' sign 
corresponds to AG~Dra, which is seen more from the pole, and 
thus its $R_{\rm h}^{\rm eff} \approx R_{\star}$ (see Sect.~2.1.). 
This implies that the other nine systems (all with a high orbital 
inclination) have $R_{\rm h}^{\rm eff} > R_{\star}$, which 
corresponds to a lower value of $X$ (see Eq.~(\ref{eqn:X})). 
To improve the results, we determined the $X(\dot M)$ 
quantities for the hot star parameters $L_{\star}, R_{\star}, 
T_{\star}$ as follows. 
(i) 
We put $T_{\star} \equiv T_{\rm h} = 160\,000$\,K, as derived 
by modelling the X-ray/UV SED of AG~Dra. 
Such a high temperature is also supported by a direct X-ray 
observation of a sample of symbiotic stars with a pronounced 
supersoft X-ray (0.1--0.4\,keV) radiation that is similar in 
the profile with that of AG~Dra 
\citep[see Fig.~2 of][]{murset+97}. 
(ii) 
We determined $R_{\star}$ and $L_{\rm H}$ for 
$L_{\star} = L_{\rm h}$. Corresponding $X(\dot M)$ values are 
plotted in Fig.~\ref{fig:mdot} by open circles. 

According to Eq.~(\ref{eqn:X}), the $X(\dot M)$ relationship 
can be written in the form 
\begin{equation}
   \log (X) = C - 2\log (\dot M), 
\label{eqn:xmd}
\end{equation}
where 
$C = \log(9.6\times 10^{-34} R_{\rm WD}^3/R_{\star}^2 L_{\rm H}) 
     + 2\log (v_{\infty})$ for 
$\alpha_{\rm B}({\rm H},20\,000$\,K) = 
1.43$\times 10^{-13}$\,cm$^{3}$s$^{-1}$ \citep[e.g.][]{humsea63} 
and given parameters. 
Fits to filled circles and to open circles with the `plus' sign 
in Fig.~\ref{fig:mdot} correspond to $C$ = -14.81 and -14.26, 
respectively (for $\dot M$ in \myr). This shows that the quantities 
of $X$ for the hot star parameters ($L_{\star}$, $R_{\star}$, 
$T_{\star}$) are a factor of $\sim 3.5$ higher than those determined 
using the observed parameters ($L_{\rm h}$, $R_{\rm h}^{\rm eff}$, 
$T_{\rm h}$). 
Thus, regardless of the selected parameters, both sets of 
the $X(\dot M)$ data are located well above the grey belt in 
the figure, i.e. without the possibility of creation of the
flared neutral disk-like zone, as predicted by the WCZ model 
(Sect.~2.4., relation (\ref{eqn:xrange})). 
In other words, the mass loss rate from the hot star during 
quiescent phases is too small to give rise to a neutral zone of 
hydrogen, because the rate of ionizing photons is sufficiently 
high to ionize even the compressed fraction of the wind. 

\subsubsection{$X(\dot M)$ relationship during active phases}

Determination of the hot star parameters during active phases 
for systems with a high orbital inclination is complicated by 
the presence of the edge-on disk, which blocks the original 
radiation from the hot star in the direction of the 
observer \citep[see Sect.~5.3.6. of][]{sk05}. 
Therefore, indirect methods based on the ratio of nebular 
line fluxes and/or the presence of the highest ionization 
states in the spectrum \citep[][]{mur+nuss94} are employed 
to estimate the hot star temperature during active phases. 
For example, using the ratio of \ion{He}{ii}\,4686\,\AA\ and
\hb\ line fluxes from the spectra of the symbiotic prototype
Z~And during its 2000-03 outburst, \citet{sok+06} derived 
$T_{\rm h} = 92\,000 \pm 20\,000$ to $160\,000 \pm 35\,000$\,K. 
During the smaller, 1997 outburst, both methods indicated 
an increase in $T_{\rm h}$ to 180\,000 K. 

More trustworthy estimates of $T_{\rm h}$ can be achieved 
only for non-eclipsing systems with the aid of 
multiwavelength modelling of the supersoft X-ray/UV SED, which 
covers the total spectrum emitted by the burning WD. An 
applicable system here is AG~Dra, the strongest supersoft 
source among classical symbiotic stars. Although the supersoft
component disappears during its active phases, the 
very high far-UV fluxes and a strong nebular continuum 
constrain $T_{\rm h}$ to be $> 150-180$\,kK, depending on 
the outburst stage \citep[see][]{sk05,sk+09}. 
As the orbital inclination of AG~Dra is low 
\citep[$i\approx 30-60^{\circ}$,][]{mika+95,ss97}, we can 
directly see the WD's pseudophotosphere and thus can 
assume that its 
$T_{\star} \approx T_{\rm h}$ and 
$R_{\star} \approx R_{\rm h}^{\rm eff}$ (see also Sect.~2.1). 
Therefore, we adopted the average hot star parameters, which were
derived from the model SEDs of AG~Dra, 
         $T_{\star} = 170\,000 \pm 7000$\,K, and 
         $R_{\star} = 0.06 \pm 0.01$\ro, 
(i.e. $L_{\star} = 2\,700 \pm 700$\lo\ and 
$L_{\rm H} = (1.4 \pm 0.4)\times 10^{47}$\,s$^{-1}$), as 
representative values for the active phases of AG~Dra, Z~And, 
AX~Per, AE~Ara, and AR~Pav. The parameters 
$v_{\infty}$ and $\dot M$ for these systems are available
in Table~1 of \cite{sk06} and Table~7 of \cite{sk+11}. 
%
The $X(\dot M)$ quantities of active symbiotic stars lie well 
within the possibility of creation of the flared neutral
disk-like zone predicted by the WCZ model (the grey belt in 
Fig.~\ref{fig:mdot}). 
In other words, during active phases the mass loss rate is so 
high that the rate of ionizing photons is not capable of 
ionizing the compressed wind around the hot star equator. 

Figure~\ref{fig:mdot} suggests that the presence or absence 
of the neutral disk-like zone in a symbiotic system during 
active or quiescent phases, respectively, is given by the 
value of the parameter $X$. During quiescent phases 
$X \approx 1$, while during active phases $X \approx 0.01$. 
This large difference results from the very different values 
of $\dot M$ in quiescent and active phases (Sect.~2.3.2.) 
and the proportionality of $X$ to $\dot M^{-2}$. 
Thus, we can conclude that creation of the neutral disk-like 
structure, indicated in the spectrum of active symbiotic 
binaries, is consistent with the WCZ model, i.e. it can be 
a result of a compression of the hot star wind, which is 
significantly enhanced during active phases. 
%
%
\begin{figure}
\centering
\begin{center}
\resizebox{\hsize}{!}{\includegraphics[angle=-90]{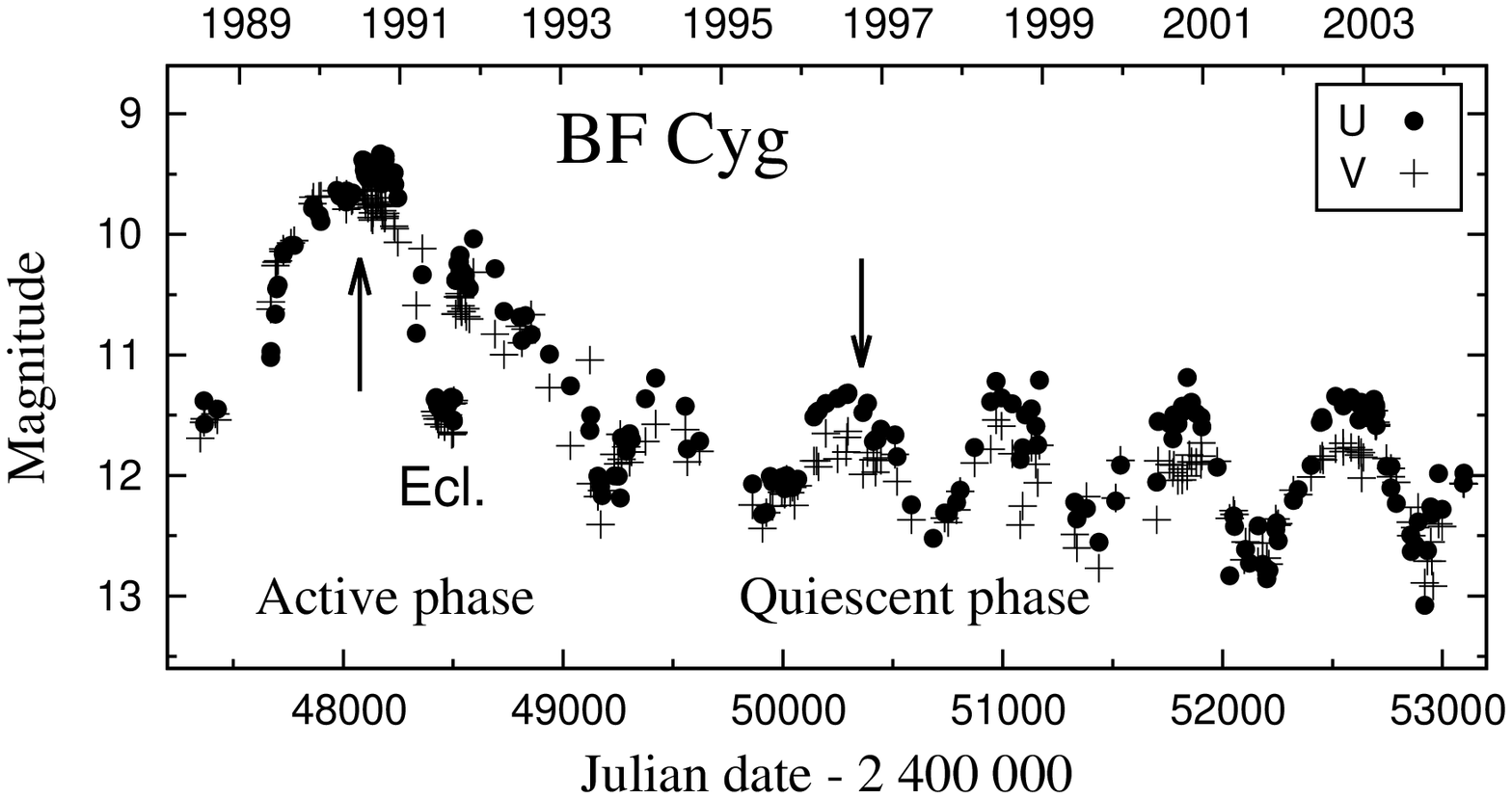}}
\resizebox{\hsize}{!}{\includegraphics[angle=-90]{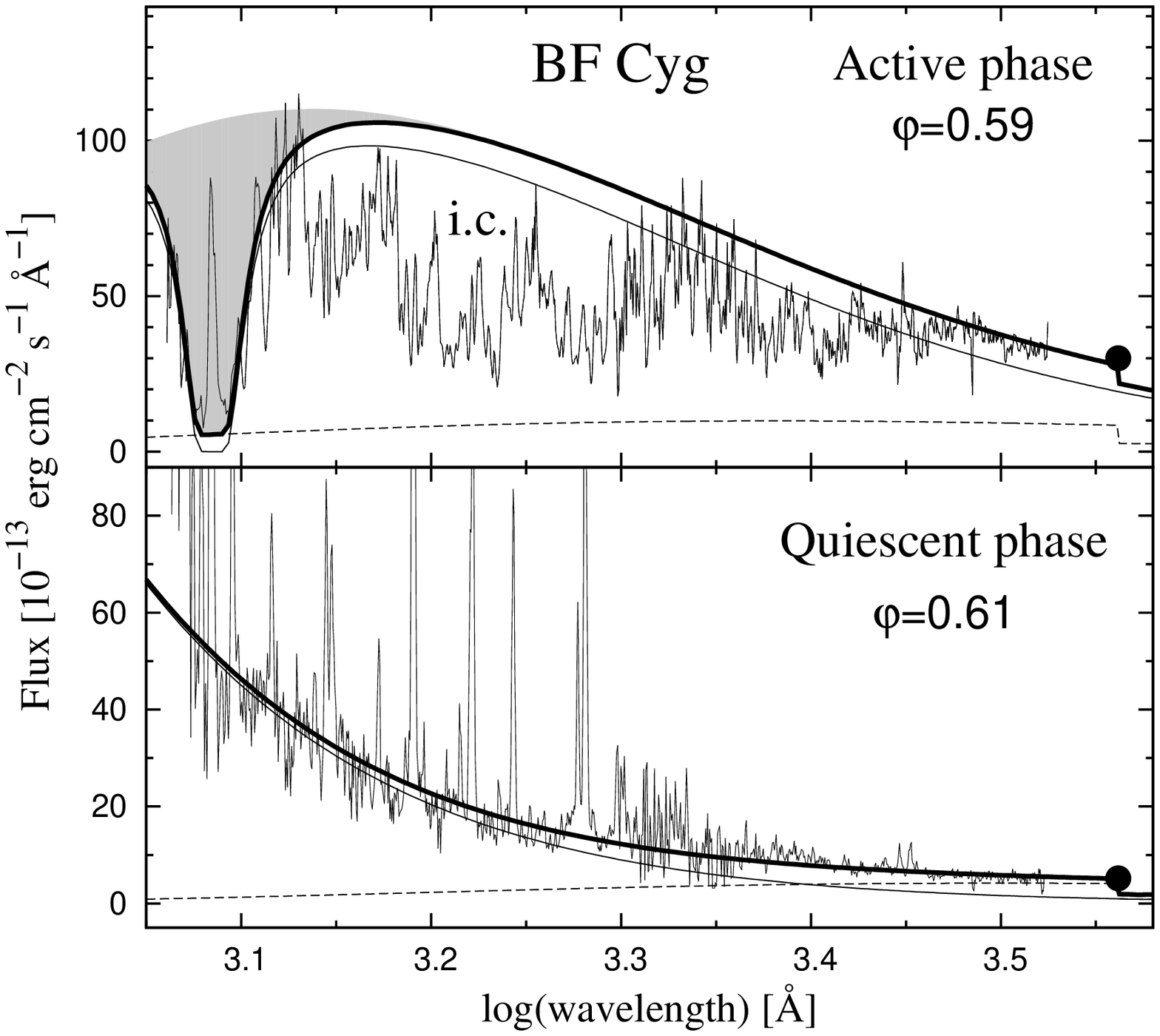}}
\end{center}
\caption[]{Ultraviolet SED of BF~Cyg during its 1990 optical 
maximum (mid) and the following quiescent phase (bottom). 
Times of observations are denoted by arrows at the top panel 
with the light curves. In both cases, the binary was seen from 
the same position with the hot star in front (orbital phase 
$\varphi \sim 0.6)$. The dashed and the thin solid line 
represent the continuum from the nebula and the hot star, 
respectively. The thick solid line is the model SED. 
The grey surface marks the Rayleigh attenuated continuum 
with $n_{\rm H} = 4.8\times 10^{22}$\cmd\ , and strong blended 
absorptions of the iron curtain are denoted by 'i.c.'.}
\label{fig:sed}
\end{figure}

\subsection{Column density of the neutral hydrogen}

In this section, we verify the model by comparing the column 
density of the neutral hydrogen atoms, which was calculated
throughout the neutral disk-like zone, with that derived from 
observations during active phases of symbiotic binaries. 

Modelling the SED of symbiotic binaries during active phases 
showed the presence of a warm ($1-2\times 10^4$\,K) disk-like 
pseudophotosphere (Sect.~1) and a large amount of the neutral 
hydrogen near to the orbital plane of the binary. Column 
densities derived from the ultraviolet spectra are between 
$\sim 10^{21}$ and a few $\times 10^{23}$ \cmd\ \citep{sk05}. 
A large amount of the neutral material can be identified even 
around the orbital phase $\sim$ 0.5, when the hot component 
is in front of the cool giant. Figure~\ref{fig:sed} demonstrates 
this case for BF~Cyg. During the active phase, the radiation 
of the warm pseudophotosphere was Rayleigh attenuated with 
$n_{\rm H} = 4.8\times 10^{22}$\cmd\ (the grey part in the 
figure), while during quiescence, the hot component radiates 
at $T_{\rm h} > 10^5$\,K with no signatures of the Rayleigh 
scattering and the iron curtain absorptions. 
This demonstrates that the presence/absence of the large 
amount of the neutral hydrogen on the line of sight is 
connected with intrinsic variations of the {\em circumstellar} 
matter around the hot star during active/quiescent phases. 
A contribution from the interstellar matter (ISM) in the 
direction to BF~Cyg can be neglected. 
For $E_{\rm B-V} = 0.35$\,mag \citep[][]{m+91}, 
$n_{\rm H}({\rm ISM}) \sim 1.7 \times 10^{21}$\cmd\ 
\citep[e.g.][]{d+s94}, which is difficult to measure on the 
low-resolution International Ultraviolet Explorer spectra.

\begin{figure}
\centering
\begin{center}
\resizebox{\hsize}{!}{\includegraphics[angle=-90]{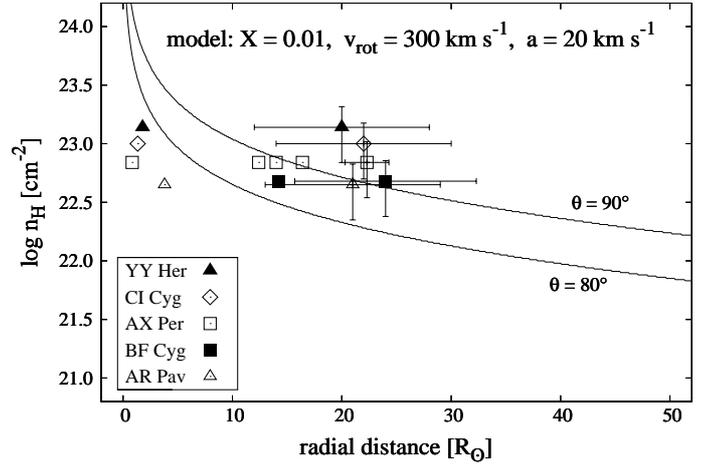}}
\end{center}
\caption[]{
Comparison of the model hydrogen column densities 
$n_{\rm H}(90^{\circ},r\ge r_{\theta})$ and 
$n_{\rm H}(80^{\circ},r\ge r_{\theta})$ (solid lines), 
calculated according to Eq.~(\ref{eqn:colden}), 
with those derived from observations of selected eclipsing 
objects, plotted at the distance $r = R_{\rm d}$ from 
the WD centre (Table~1).
}
\label{fig:nh}
\end{figure}

In the WCZ model, we calculated the column density of the neutral 
hydrogen atoms in the radial direction $\theta$ throughout the 
neutral zone as a function of $r$, 
\begin{equation}
  n_{\rm H}(\theta,r\ge r_{\theta}) = 
            \int_{\rm r\ge r_{\theta}}^{\infty}
                     N_{\rm H}(r,\theta)\,{\rm d}r,
\label{eqn:colden}
\end{equation}
where the density distribution in the wind, $N_{\rm H}(r,\theta)$, 
is given by Eq.~(\ref{eqn:nh}). 
Figure~\ref{fig:nh} shows two $n_{\rm H}(\theta,r\ge r_{\theta})$ 
functions corresponding to ionization boundaries calculated 
for $X = 0.01$, $v_{\rm rot} = 300$\kms, $a = 20$\kms\ 
(see the top right panel in Fig.~\ref{fig:h1h2}), and two different 
orbital inclinations $i = \theta = 80^{\circ}$ and 90$^{\circ}$. 
The edge-on view ($\theta = 90^{\circ}$) gives a maximum value 
of the column density, since we are looking through the densest 
parts of the compressed wind. 
Calculations were performed for $T_{\star} = 1.7\times 10^{5}$\,K 
and $L_{\star} = 3\times 10^{3}$\lo, 
i.e. $L_{\rm H} = 1.6\times 10^{47}$\,s$^{-1}$ (Sect.~3.1.2.), 
and the wind parameters, 
$\dot{M} = 2\times10^{-6}$\myr, $v_{\infty} = 2\,000$\kms, 
which represent mean values measured during active phases 
\citep[see Table~1 of][]{sk06}. 

To compare the observed values of $n_{\rm H}$ with the model 
(i.e. $n_{\rm H}(\theta,r=R_{\rm d})$, 
see Eq.~(\ref{eqn:colden})), we need to know the radius of 
the optically thick disk, $R_{\rm d}$, whose outer rim represents 
the warm pseudophotosphere indicated during active phases of 
symbiotic binaries with a high orbital inclination 
\citep[Sect.~5.3.5 and Fig.~27 in][]{sk05}. 
Therefore, a rough estimate of $R_{\rm d}$ can be obtained from 
the effective radius, $R_{\rm h}^{\rm eff}$, of the hot components 
derived from the SED. According to its definition 
(here in Sect.~2.1.), we can write 
\begin{equation}
  4\pi (R_{\rm h}^{\rm eff})^2\sigma T_{\rm h}^4 = 
  2\pi R_{\rm d} 2 H \sigma T_{\rm h}^4, 
\label{eqn:reff}
\end{equation}
where $H$ is the height of the flared disk 
(see also Fig.~\ref{fig:model}). 
For a canonical value of $H/R_{\rm d} \equiv 0.3$ during active 
phases \citep[][]{sk06}, $R_{\rm d}\sim 2\times R_{\rm h}^{\rm eff}$. 
Table~1 summarizes values of $R_{\rm h}^{\rm eff}$ from Table~4 of 
\cite{sk05} and Table~8 of \cite{sk+11}. 
%
%
\begin{table}
\begin{center}
\caption{Column densities $n_{\rm H}$, effective radii of 
         the hot components $R_{\rm h}^{\rm eff}$, 
         disk radii $R_{\rm d}$, and emission measures $EM$ 
         derived from observations. 
         Data are from \cite{sk05} and \cite{sk+11}.}
\begin{tabular}{lccccc}
\hline
\hline\\[-3mm]
 Object & $\log\,n_{\rm H}$ & $R_{\rm h}^{\rm eff}$ & $R_{\rm d}^{1)}$ 
        & $R_{\rm d}^{2)}$  & $EM$ \\
        & (\cmd)            & (\ro) & (\ro) & (\ro) 
        & $(10^{59}$\cmt) \\
\hline\\[-2mm]
 BF~Cyg & 22.68$^{+0.18}_{-0.30}$ & 7.1  & 13.0 & 24$\pm 8.3$ 
        & $ < 120 $ \\[1mm]
 CI~Cyg & 23.00$^{+0.18}_{-0.30}$ & 0.67 & 1.23 & 22$\pm 8.0$ 
        & $ 5.3 $ \\[1mm]
 YY~Her & 23.14$^{+0.18}_{-0.30}$ & 0.89 & 1.63 & 20$\pm 8.0$
        & $ 7.2 $ \\[1mm]
 AR~Pav & 22.65$^{+0.18}_{-0.30}$ & 1.90 & 3.48 & 21$\pm 8.0$
        & $ 17 $ \\[1mm]
 AX~Per & 22.84$^{+0.18}_{-0.30}$ & 0.42 & 0.77 & ---
        & 3.6 \\[1mm]
        &  ---                    & 6.2  & 11.3 & ---
        & 3.5 \\[1mm]
        &  ---                    & 8.2  & 15.0 &  --- 
        & 3.8 \\[1mm]
        &  ---                    & 11   & 20.1 & 22$\pm 2.0$ 
        & 8.2 \\[1mm] 
        &  ---                    & 7.0  & 12.8 &  --- 
        & 4.0 \\[1mm]
 AG~Dra &  ---                    & ---  & ---  & ---
        & 4.4 \\
\hline
\end{tabular}
\end{center}
$^{1)}$ according to Eq.~(\ref{eqn:reff}),~ 
$^{2)}$ from the eclipse profile (see text). 
\normalsize
\end{table}
%

The disk radius can also be estimated directly from eclipses, 
which develop in optical light curves during active phases 
\citep[e.g.][ Fig.~4 here]{bel91,munari92,brandi+05,sk08}. 
For the 1991 eclipse of BF~Cyg, the radius of the eclipsed object 
in units of the binary components separation, $A$, was 
$R_{\rm e}/A = 0.051 \pm 0.018$, which yields 
$R_{\rm e} (\equiv R_{\rm d}) = 24 \pm 8.3$\ro\ for 
the total mass of the binary $M_{\rm T} = 2.5$\mo\ and 
the orbital period of 757.3 days 
\citep[][ and references therein]{sk+97}. 
CI~Cyg: The first two contact times of its 1975 eclipse, 
 $T_1 = JD\,2\,442\,634.5 \pm 1.5$, 
 $T_2 = JD\,2\,442\,646.8 \pm 2.0$ 
\cite[from the light curve of][]{bel91}
and its middle, 
$JD_{\rm Ecl.}(1975.8) = 2\,442\,691.6 \pm 0.8$ 
\citep[][]{sk98}, 
orbital period of 853.8 days \citep[][]{fekel+00} and 
$M_{\rm T} = 2.0$\mo\ \citep[][]{kenyon+91} give 
$R_{\rm d} = 22 \pm 8$\ro. 
YY~Her: The contact times of the eclipse that developed during 
the 1993-94 active phase \citep[][]{sk05}, the orbital period 
of 574.6 days \citep[][]{wiecek+10} for $M_{\rm T} \equiv 2.0$\mo\ 
gives $A = 367$\ro\ and thus $R_{\rm d} = 20 \pm 8$\ro.
AR~Pav: During epochs 59 -- 63 (1987 -- 1995) of its historical 
light curve, \cite{sk+00} estimated $R_{\rm e}/A = 0.05 \pm 0.02$ 
(see Fig.~1 there), which for $A = 419$\ro\ \citep[][]{schild+01} 
implies $R_{\rm d} = 21 \pm 8$\ro. 
AX~Per: Recently, \citet{sk+11} derived $R_{\rm d} = 27 \pm 2$\ro\ 
from the 2009 eclipse, which developed during the 2007-10 active 
phase. The eclipse profile was asymmetric, with the first two 
contact times corresponding to $R_{\rm d} = 22.3\pm 2$\,\ro, 
in particular agreement with the disk radius derived from 
the model SED ($R_{\rm h}^{\rm eff} = 11$\ro\ and
Eq.~(\ref{eqn:reff})). 
It is important to note that the radius of the warm disk-like 
shell, derived from the eclipse profile, represents an upper 
limit because of a measurable contribution of the ionized part 
of the hot star wind at radial distances of $\sim 20$\ro\ 
($EM(r > 20\,R_{\sun}) \sim 10^{57}$\cmt\ in the model). This 
also causes large uncertainties of $R_{\rm d}$ from eclipses. 
%
%

Figure~\ref{fig:nh} compares quantities of $n_{\rm H}$ 
derived from observations of symbiotic stars with a high orbital 
inclination, BF~Cyg, CI~Cyg, YY~Her, AR~Pav, and AX~Per.
We used column densities derived from the Rayleigh scattering of 
the far-UV continuum photons around the Ly-${\alpha}$ line on 
the neutral atoms of hydrogen. 
The errors in $n_{\rm H}$ this way derived are of 
the order of 50\% because of the influence of other absorption 
effects \citep[][]{d+99}. 
For the purpose of this work, we used values of $n_{\rm H}$ 
for the above-mentioned symbiotic stars during their outbursts 
at the orbital phases between 0.3 and 0.6 to eliminate a larger 
influence of the neutral wind from the giant (Table~1). 
Nevertheless, the measured values of $n_{\rm H}$ can still 
contain a contribution from the giant's wind, which is in the 
neutral form at/around the orbital plane during active phases, 
because the ionizing photons are blocked by the neutral disk-like 
formation surrounding the hot star. For the mass loss rate from 
the giant of a few $\times 10^{-7}$\,\myr\ \citep[][]{sk05}, 
the column density of its wind in the direction to the hot 
component at the orbital phase 0.5 and $i = 90^{\circ}$ 
is of $\approx 1\times 10^{22}$\cmd\ \citep[][]{sek+skop}.

In spite of large errors in $n_{\rm H}$ and $R_{\rm d}$, often 
derived from non-simultaneous observations, Fig.~\ref{fig:nh} 
shows that the $n_{\rm H}$ values are consistent with those 
calculated within the WCZ model for parameters of the hot star 
and its wind derived from observations during active phases. 

\subsection{Emission measure of the ionized wind from the hot star}

In this section we calculate the emission measure, $EM$, of 
the ionized part of the hot star wind from regions above/below 
the neutral disk-like zone and compare it to that derived from 
observations during active phases.

Modelling the SED of symbiotic binaries during active phases 
showed that the emission measure of the nebular component of 
radiation is between a few $\times 10^{58}$ and a few 
$\times 10^{60}$ \cmt\ \citep[see Table~4 of][]{sk05}. 
In addition, the modelling revealed the presence of a 
low-temperature nebula (LTN, $T_{\rm e} \sim 14\,000$\,K), 
which is subject to eclipses, and a high-temperature nebula 
(HTN, $T_{\rm e} > 30\,000$\,K), which is seen during eclipses 
as the only component. 
Later, \cite{sk06} ascribed the LTN to the ionized wind from 
the hot star, which develops during active phases. 
%
\begin{figure}
\centering
\begin{center}
\resizebox{\hsize}{!}{\includegraphics[angle=-90]{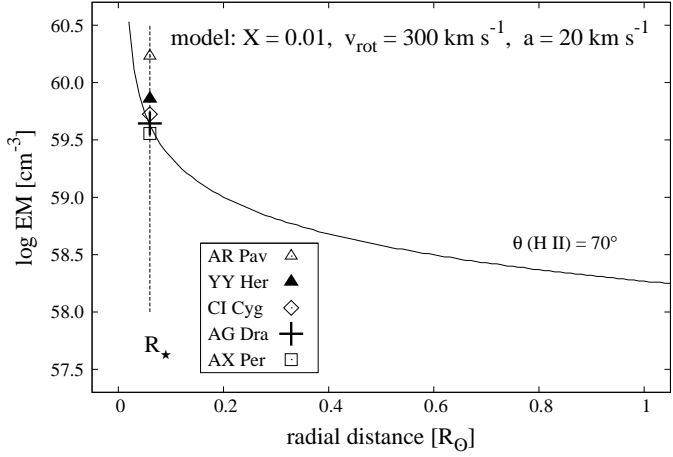}}
\end{center}
\caption[]{
Comparison of the model emission measure calculated according 
to Eq.~(\ref{eqn:em2}) throughout the \ion{H}{ii} region as 
a function of the radial distance $r$ (solid line) with 
the values derived from the spectra of selected symbiotic 
stars during active phases (Table~1). $\theta$(\ion{H}{ii}) 
is the opening angle of the \ion{H}{ii} zone. The same model 
as in Fig.~\ref{fig:nh} was selected. 
            }
\label{fig:em}
\end{figure}
%
In our model, $EM$ is determined as
\begin{equation}
  EM = \int_{V} N_{\rm H}^{2} (r,\theta) \,{\rm d}V,
\label{eqn:em}
\end{equation}
where $N_{\rm H}(r,\theta)$ is the density of the ionized
hydrogen given by Eq.~(\ref{eqn:nh}), calculated throughout 
the volume $V$ of the \ion{H}{ii} zone. In the spherical 
coordinates ($r, \theta, \phi$), the volume element 
${\rm d} V = 
r^{2} \sin \theta \,{\rm d}r\,{\rm d}\theta\,{\rm d}\phi$, 
and in the WCZ model, the density distribution is azimuthally 
symmetric (i.e. along the coordinate $\phi$). This allows us 
to express Eq.~(\ref{eqn:em}) as 
\begin{equation}
  EM = 4\pi \int_{\,R_{\star}}^{\,\infty} 
       \int_{\,0}^{\,\theta(\ion{H}{ii})} N_{\rm H}^{2}(r,\theta)\,
       r^{2}\sin \theta\,{\rm d}r\,{\rm d}\theta,
\label{eqn:em2}
\end{equation}
where $\theta(\ion{H}{ii})$ is the $\theta$ coordinate of the 
\ion{H}{i}/\ion{H}{ii} boundary at the given $r$. 
Figure~\ref{fig:em} shows the calculated $EM$ of the ionized 
part of the wind as a function of $r$ for the same model 
that we used to calculate $n_{\rm H}$ in Sect.~3.2. 
It shows that the observed values of the emission measure, 
as derived by modelling the SED 
\citep[i.e. $EM_{\rm LTN}$ in Table~4 of][ Table~1 here]{sk05}, 
are in a good agreement with the model prediction. 
We excluded the case of BF~Cyg, because it was not possible to 
extract the LTN component from the SED. 
A problem in comparing the observed and modelled emission 
measure is connected with a shielding effect for objects with 
a high orbital inclination. In the real case, it is probable 
that the outer rim of the flared disk will shield the inner 
parts of the ionized wind. As a result, the calculated emission 
measure $EM_{\rm cal} > EM_{\rm LTN}$. 
However, we did not include the shielding effect in 
the calculation, because the orbital inclination and 
the opening angle of the neutral zone are not well-known 
parameters.

According to Fig.~\ref{fig:em}, we can conclude that the 
emission measure of the ionized wind from the hot star, 
calculated within the WCZ model (Eq.~(\ref{eqn:em2})),
is consistent with that derived from observations by
modelling the SED of symbiotic binaries during active phases. 
%
%
\begin{figure}
\centering
\begin{center}
\resizebox{\hsize}{!}{\includegraphics[angle=-90]{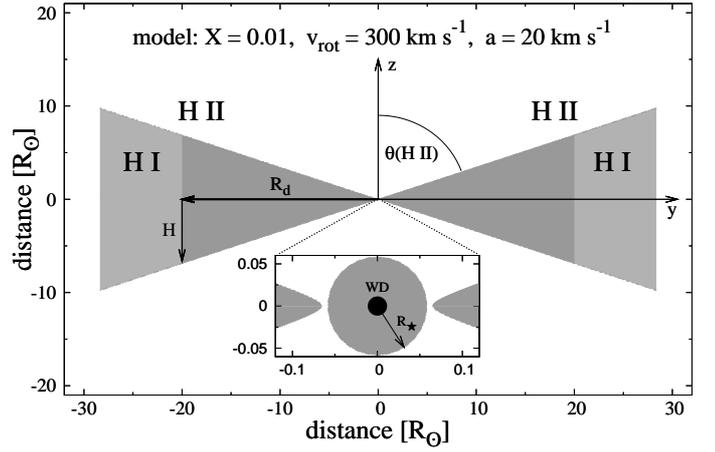}}
\end{center}
\caption[]{
Model of the ionization structure of the hot components in 
symbiotic binaries during active phases (Sect.~2, 
Eq.~\eqref{eqn:calcX}) 
as seen on a cut perpendicular to the orbital ($x,y$) plane 
containing the WD, which spins along the $z$-axis. 
The neutral zone has the form of a flared disk. It is optically 
thick to the distance $R_{\rm d}$, i.e. the radius of the warm 
pseudophotosphere. The ionized zone has the opening angle 
$\theta$(\ion{H}{ii}). The model corresponds to 
$R_{\rm h}^{\rm eff} \sim 11$\ro\ (see Eq.~\eqref{eqn:reff}). 
The inplot figures the central WD as a black circle with the 
radius $R_{\rm WD}$, which is the source of the optically thick 
wind. The wind becomes optically thin at the distance $R_{\star}$ 
(the shadow circle), which simulates the WD's 
pseudophotosphere, the source of ionizing photons. 
          }
\label{fig:model}
\end{figure}

\section{Conclusion}

We investigated the ionization structure around the hot stars 
in symbiotic binaries during active phases. Applying the wind 
compression model (Sect.~2), we found that a neutral disk-like 
zone can be formed as a result of the enhanced wind from the 
rotating WD at its equatorial plane. 
Around the pole directions, above/below the neutral disk, the 
wind is diluted by a fast rotation, which makes it easily 
ionized by the central hot star (i.e. the WD's pseudophosphere 
in the model). Examples of the corresponding 
\ion{H}{i}/\ion{H}{ii} boundaries are shown in 
Fig.~\ref{fig:h1h2}. 
A representative model of the ionization structure of the hot 
component in a symbiotic binary during active phase is shown 
in Fig.~\ref{fig:model}. 

Applying the wind compression model to physical parameters of 
the hot stars in symbiotic binaries, we found that the neutral 
disk-like zone can be created during active phases, when the 
mass loss rate enhances to $\sim 2\times 10^{-6}$\myr. However,
during quiescent phases, the formation of the neutral region 
is unlikely because of insufficient mass loss rate only of a few 
$\times 10^{-8}$\,\myr\ (Sect.~3.1., Fig.~\ref{fig:mdot}). 

We verified the model by comparing the hydrogen column density, 
calculated in the radial direction from the hot star throughout 
the \ion{H}{i} zone and the emission measure of the \ion{H}{ii} 
zone, with quantities derived from observations 
(Sects.~3.2. and 3.3.). Figs.~\ref{fig:nh} and \ref{fig:em} show 
that both parameters are consistent with those calculated 
according to the wind compression model for
$\dot{M} = 2\times10^{-6}$\myr\ and $v_{\infty} = 2\,000$\kms, 
which represent mean values derived from observations 
during active phases.

In view of these results, we propose that the flared 
optically thick disks, indicated in the spectrum of active 
symbiotic binaries with a high orbital inclination, can form 
from the enhanced wind, which is compressed towards the 
equatorial plane due to a fast rotation of its source, the 
central WD. In this way we justified the ionization structure 
of the hot components in symbiotic binaries during active 
phases, as we had suggested in previous papers 
\citep[e.g.][]{sk05,sk06,sk+06,sk+11}. 

\begin{acknowledgement}
The authors thank the anonymous referee for constructive 
comments. 
This research was supported by a grant of the Slovak Academy of
Sciences, VEGA No. 2/0038/10 and by the Project ITMS No. 26220120029,
based on the supporting operational Research and development
program financed from the European Regional Development Fund.
\end{acknowledgement}

\end{document}